\documentstyle[12pt]{article}\textheight 220mm\textwidth 160mm
\newcommand{\en}[1]{(\ref{#1})} 

\newcommand{\al}{\alpha}	\newcommand{\be}{\beta}
\newcommand{\g}{\gamma}		\newcommand{\de}{\delta}
\newcommand{\e}{\epsilon}	\newcommand{\ve}{\varepsilon}
\newcommand{\z}{\zeta}		\newcommand{\th}{\theta}
	
\newcommand{\k}{\kappa}		\newcommand{\lam}{\lambda}
\newcommand{\m}{\mu}		\newcommand{\n}{\nu}
\newcommand{\x}{\xi}		\newcommand{\p}{\pi}
\newcommand{\r}{\rho}		\newcommand{\s}{\sigma}
		\newcommand{\f}{\phi}
\newcommand{\vp}{\varphi}	\newcommand{\w}{\omega}
\newcommand{\h}{\eta}

\newcommand{\G}{\Gamma}		
\newcommand{\C}{\Theta}		\newcommand{\Lam}{\Lambda}

\newcommand{\F}{\Phi}		\newcommand{\V}{\Psi}
\newcommand{\W}{\Omega}		

\newcommand{\CC}{{\cal C}}

\newcommand{\CJ}{{\cal J}}

\newcommand{\CP}{{\cal P}}

\newcommand{\diag}{{\rm diag}}

\newcommand{\dl}{\partial}

\newcommand{\bn}{\begin{eqnarray}}
\renewcommand{\en}{\end{eqnarray}}
\newcommand{\nn}{\nonumber}

\begin{document}

\hspace*{11cm}YITP-U-95-5\\
\hspace*{11.6cm}February 1995\\
\begin{center}
{\Large\bf BFV-BRST Quantization of 2D Supergravity}
\end{center}

\vspace*{.5cm}
\def\thefootnote{\alph{footnote}}
\begin{center} {\sc T. Fujiwara}$^{1}$
, {\sc Y. Igarashi}$^{2,}$
\footnote{Permanent address: Faculty of Education, Niigata University,
Niigata, 950-21, Japan.}
, {\sc R. Kuriki}$^{3}$
and  {\sc T. Tabei}$^{4,}$
\end{center}
\vspace*{0.2cm}
\begin{center}
{\em $\ ^{1}$ Department of Physics, Ibaraki University,
Mito 310, Japan}\\
{\em $\ ^{2}$ Yukawa Institute for Theoretical Physics, Kyoto
University, Uji 611, Japan}\\
{\em $\ ^{3}$ Department of Physics, Tokyo Institute of Technology,
Tokyo 152, Japan}\\
{\em $\ ^{4}$ Graduate School of Science and Technology,
Kanazawa University, \\
Kanazawa 920-11, Japan}
\end{center}
\vfill
\begin{center}
{\large\sc Abstract}
\end{center}
\noindent
Two-dimensional supergravity theory is quantized as an anomalous gauge
theory.
In the Batalin-Fradkin (BF) formalism, the anomaly-canceling
super-Liouville fields are introduced to identify the original
second-class constrained system with a gauge-fixed version of a
first-class system. The BFV-BRST quantization applies
to formulate the theory in the most general class of gauges.
A local effective action constructed in the configuration space
contains two super-Liouville actions;
one is a noncovariant but
local functional written only in terms of 2D supergravity fields,
and the other contains the super-Liouville fields canceling the
super-Weyl anomaly.
Auxiliary fields for the Liouville and
the gravity super-multiplets are introduced to make the
BRST algebra close off-shell.
Inclusion of them turns out to be essentially important especially in
the super-lightcone gauge-fixing, where the super-curvature equations (
$\dl^3_-g_{++}=\dl^2_-\chi_{++}=0$) are obtained as a result of BRST
invariance of the theory.  Our approach reveals the
origin of the graded-SL(2,R) current algebra symmetry in a transparent
manner.

\def\thefootnote{\fnsymbol{footnote}}

\newpage
\pagestyle{plain}

\section{Introduction}
\setcounter{equation}{0}

During the last decade there has been remarkable progress
in our understanding of noncritical string theories \cite{pol2,ct,m}. The
first key observation of Polyakov \cite{pol2} was that
the conformal mode of the metric variables does not decouple
from the theory at noncritical dimensions. Along the line
of thought, noncritical string has investigated in the
light-cone gauge \cite{pol1,kpz}. Noting the SL(2,R)
Kac-Moody symmetry, Knizhnik, Polyakov and Zamolodchikov
(KPZ) have succeeded in deriving gravitational scaling
dimensions for conformal matters interacting with the 2D
gravity on the world-sheet. Furthermore, David  \cite{d},
Distler and Kawai (DDK) \cite{dk} showed that the Polyakov's path
integral formulation \cite{pol2} reproduces the KPZ's results
also in the conformal gauge.
It was based on the assumption that the Jacobian associated with
changing the functional measure from that for the conformal
mode defined in \cite{pol2} to translational invariant
one generates a Liouville type action.
This DDK ansatz for the functional measure has been
examined using the heat kernel method \cite{mmh}.
It has also been discussed the relation between the
conformal gauge and the light-cone gauge \cite{gn,fikt},
and analyses based on the
BRST formalism have been carried out
by several authors \cite{kura,ih}.
Furthermore, the analysis
in the light-cone gauge and conformal gauge have been
extended to supersymmetric case in
\cite{gx,pz,aaz} and in \cite{dk,dhk}, respectively.

In our previous paper \cite{ftimk},
we gave a systematic
canonical formulation of Polyakov string at noncritical dimensions
by applying the idea developed for anomalous gauge theory
\cite{fik1}. It provides a general approach to noncritical
strings. The BRST anomalies of the Polyakov string theory at
noncritical dimensions \cite{ko,FujG,fikm} can be compensated by
introducing new degrees of freedom and, thereby, the theory can
be made gauge symmetric, i.e., invariant under Weyl
rescalings of the metric variables as well as world-sheet
reparametrizations. The BRST gauge-fixed action turned out to
contain two Liouville type actions, the one being written only
in world-sheet metric and the other containing the new degree
as Liouville field. In the conformal gauge, this reduces to
the effective action of DDK, giving a justification of their
functional measure ansatz from canonical view point. We further
gave a systematic description of the theory in the light-cone gauge,
clarifying the relation between the BRST invariance and SL(2,R)
Kac-Moody symmetry.

In this paper we will investigate the extension of this work to
Nevue-Schwarz-Ramond superstring \cite{ns,imnu}. The locally
supersymmetric action \cite{dz} can be regarded as N=1 2D supergravity
(SUGRA) coupled with string variables as superconformal matters. The
basic strategy parallels the bosonic string case. Our starting point
is the most general form of the BRST anomalies in 2D SUGRA
\cite{fikkt}
in the extended phase space (EPS) of Batalin, Fradkin and Vilkovisky
(BFV) \cite{bfv}. The anomaly appeared there as anomalous Schwinger
terms which destroy first-class nature of the super-Virasoro
constraints. The quantum system is described by the second-class
constraints. In general, systems with second-class constraints can be
regarded as gauge-fixed systems of underlying symmetric theories.
Actually, one can rewrite the system with second-class constraints
into a gauge symmetric form by introducing compensating
fields in EPS.
Batalin and Fradkin (BF) \cite{bf} developed a general idea
of converting systems with second-class constraints into
gauge symmetric ones in a general and systematic way. Applying their
method to the present case and carrying out the BFV-BRST quantization
\cite{bfv}, one obtains a BRST gauge-fixed effective action.
The construction of the action given
here completely parallels that in the anomalous chiral gauge
theories \cite{fik1,fik2}. The resulting effective action contains two
super-Liouville actions; One just coincides with the local but
noncovariant counterterm found in \cite{fikkt}, and acts as the
Wess-Zumino-Witten term to shift the super-Virasoro anomaly to
the super-Weyl anomaly. The other with the BF variables as
super-Liouville fields cancels this super-Weyl anomaly. This can
be shown without invoking a particular gauge. The fact that our
action reduces exactly to the one suggested in
\cite{dk,dhk} for the superconformal gauge implies that our
formulation provides a justification of their
functional measure ansatz from canonical view point.

Although our construction of the effective action almost parallels that
for the bosonic theory, some new issues arise in the quantization
of the fermionic theory. In the EPS, BRST transformation
incorporates supersymmetry transformation, and closes off-shell by
construction. One may go to the configuration space by eliminating
the momentum variables. In general, the BRST transformation in the
configuration space thus obtained closes only on-shell. In the
superconformal gauge, the on-shell closure of the algebra is enough to
quantize the theory. This is not the case, however, in the supersymmetric
light-cone gauge. Inclusion of auxiliary fields for the supermultiplets
of the gravity and the Liouville sectors needed to close the algebra off-shell
turns out to be essentially important for quantization.
 Since systematic way of introducing
such fields in EPS is not known, we shall discuss the auxiliary field
as well as the complete form of the BRST transformation after passing
to the configuration space. It should be noted in our general construction
that the auxiliary field for the Liouville sector and that for the gravity
sector couple non-trivially, generating a new local symmetry
incorporated in the final form of the BRST transformation.

This paper is organized as follows. In Section 2, we briefly outline
the most general form of BRST anomaly in 2D SUGRA. The BF algorithm is
applied to cancel the super-Virasoro anomaly. We formulate in Section 3
the BRST gauge-fixed effective action in the EPS and describe
covariantization of the action, where auxiliary fields are introduced
to obtain the off-shell nilpotent BRST transformations in the
configuration space. The superconformal gauge is discussed in Section 4.
Section 5 is devoted to supersymmetric light-cone gauge-fixing.
OSp(1,2) Kac-Moody symmetry is obtained in a systematic manner
based on the BRST invariance.  Summary and discussion is given in
Section 6. In Appendix, we summarize the BRST transformation
in the configuration space.

\section{BRST anomaly in the fermionic string theory}
\setcounter{equation}{0}

In this section we will briefly review the BFV formalism for
fermionic string theory and the BRST anomaly along the line
of ref. \cite{fikkt}. The fermionic string can be formulated
as $N=1$ 2D SUGRA coupled to string coordinates and
is described by the action \cite{dz}
\begin{equation}
S_X=-\int d^2xe\biggl[~\frac{1}{2}
\bigl(g^{\al\be}\dl_\al X\dl_\be X
-i\overline\psi\r^\al\nabla_\al\psi\bigr)
+\overline\chi_\al\r^\be\r^\al\psi\dl_\be X
+\frac{1}{4}\overline\psi\psi\overline\chi_\al\r^\be\r^\al
\chi_\be~\biggr]~,
\label{str}
\end{equation}
where $X^\m$ and $\psi^\m$ ($\m=0,\cdots,D-1$) are, respectively,
the bosonic and fermionic string variables,\footnote{We choose
$\h^{ab}=\diag(-1,1)$ and $\h^{\mu\nu}=\diag(-1,1,\cdots,1)$ for
flat metrics and $\e^{ab}=-\e^{ba}$ with $\e^{01}=1$ for Levi-Civita
symbol. The world-sheet coordinates are denoted by $x^\al
=(\tau,\s)$ for $\al=0,1$ and are assumed to take $-\infty<\s<+\infty$.
It is straightforward to make the analysis on a finite interval of
$\s$ so as to impose the Neveu-Schwarz or Ramond boundary conditions.
We will use the notation $\dot A=\dl_\tau A$ and $A'=\dl_\s A$ for
derivatives. Dirac matrices $\r^a$ ($a=0,1$) are chosen to be
$\r^0=\s_2$, $\r^1=i\s_1$, and $\r^5\equiv\r^0\r^1=\s_3$, where
$\s_k$ ($k=1,2,3$) are Pauli matrices.} and we have suppressed the
space-time indices. The zweibein and gravitino are denoted by
$e_\al{}^a$ and $\chi_\al$, respectively.

The action (\ref{str}) possesses invariances under the world-sheet
reparametrizations, local Lorentz rotations, local supersymmetry,
local Weyl rescalings and fermionic symmetry \cite{dz}. This suggests a
convenient choice of parametrizations \cite{fikkt} for the zweibein and
the gravitino as
\begin{eqnarray}
&&\lam^\pm=\pm\frac{e_0{}^\pm}{e_1{}^\pm}
=\frac{\sqrt{-g}\pm g_{01}}{g_{11}}
{}~, \qquad\x=\ln(-
e_1{}^+e_1{}^-)=\ln g_{11}~,
\qquad \ve=\frac{1}{2}\ln\biggl(-\frac{e_1{}^+}{e_1{}^-}\biggr)~, \nn \\
&&\nu_\pm=\frac{(\chi_0\pm\lam^\mp\chi_1)_\pm}{\sqrt{\mp e_1{}^\mp}}~,
\qquad \Lam_\pm=\frac{4\chi_{1\mp}}{\sqrt{\pm e_1^\pm}}~,
\label{zweibein}
\end{eqnarray}
where $\chi_{\al\mp}$ stands for the upper and lower components of
$\chi_\al$, and $e_\al{}^\pm=e_\al{}^0\pm e_\al{}^1$.
We also use
the rescaled components $\psi_\pm$ defined by
\begin{equation}
\psi=\pmatrix{e^{\ve-\frac{\x}{2}}\psi_-\cr
e^{-\ve-\frac{\x}{2}}\psi_+}
\label{fermion}
\end{equation}
for the fermionic string variables. In this parameterization $\xi$,
$\ve$ and $\Lam_\pm$ are the only variables that change
under the Weyl rescalings, the Lorentz rotations and the fermionic
symmetry, respectively.
The other variables are all invariant under these symmetries,
and the action (\ref{str}) can be written only in terms of these variables as
\begin{eqnarray}
S_X&=&\int d^2x\biggl[~\frac{1}{\lam^++\lam^-}
(\dot X-\lam^+X')
(\dot X+\lam^-X') \nonumber \\
&&~~~~~~~~~+\frac{i}{2}\psi_+(\dot\psi_+-\lam^+\psi_+')
+\frac{i}{2}\psi_-(\dot\psi_-+\lam^-\psi_-') \nonumber\\
&&~~~~~~~~~+\frac{2}{\lam^++\lam^-}\bigl\{i(\dot X-\lam^+ X')\psi_
-\n_+-i(\dot X+\lam^-X')\psi_+\n_-\bigr\} \nonumber\\
&&~~~~~~~~~+\frac{2}{\lam^++\lam^-}\psi_+\psi_-\n_+\n_- ~\biggr]~.
\label{stract}
\end{eqnarray}

In the canonical description the local symmetries manifest
themselves as first-class constraints according to Dirac's
classification \cite{DirA}. Denoting the canonical momenta for $X^{\m}$,
$\lam^\pm$, $\x$, $\ve$, $\n_\pm$ and $\Lam_\pm$ by $P_{\m}$,
$\p^\lam_\pm$, $\p_\x$, $\p_\ve$, $\p_\n^\pm$ and $\p_\Lam^\pm$,
respectively, the canonical theory of this system is characterized
by the following set of first-class constraints
\begin{eqnarray}
\vp_{\hat A}&\equiv&\p_{\hat A}\approx0
\qquad {\rm for~~}\hat A=\lam^\pm,\x,\ve \nonumber \\
{\cal J}^{\hat z}&\equiv& \p^{\hat z}\approx0 \qquad {\rm for}
{}~~\hat z=\n_\pm,\Lam_\pm~, \nonumber \\
\vp^X_\pm&\equiv&{1\over4}(P\pm X')^2\pm{i\over2}\psi_\pm\psi_\pm'\approx0~,
\nonumber \\
\CJ^X_\pm&\equiv&\psi_\pm(P\pm X')\approx0~,
\label{sVconst}
\end{eqnarray}
where $\vp^X_\pm$ and $\CJ^X_\pm$ are the super-Virasoro constraints,
and satisfy the classical super-Virasoro algebra under the equal-$\tau$
super-Poisson brackets \cite{bfv}
\begin{eqnarray}
&&\{\vp^X_\pm(\s),\vp^X_\pm(\s')\}
=\pm(\vp^X(\s)+\vp^X(\s'))\dl_\s\de(\s-\s')~, \nonumber \\
&&\{\CJ^X_\pm(\s),\vp^X_\pm(\s')\}
=\pm{3\over2}{\cal J}^X_\pm(\s)\dl_\s\de(\s-\s')
\pm\dl_\s\CJ^X_\pm\de(\s-\s')~, \nn \\
&&\{\CJ^X_\pm(\s),\CJ^X_\pm(\s')\}=-4i\vp^X_\pm(\s)\de(\s-\s')~,\nn \\
&&{\rm all~other~super\hbox{-}Poisson~brackets~vanish.} \label{csValg}
\end{eqnarray}

To set up the BFV-BRST formalism \cite{bfv}, we introduce the EPS
by adding to the classical phase space the
ghost-auxiliary field sector for each constraint as
\begin{eqnarray}
\vp_{\rm A}:&&({\cal C}^{\rm A},\overline{\cal P}_{\rm A}),\quad
({\cal P}^{\rm A},\overline{\cal C}_{\rm A}),\quad
(N^{\rm A},B_{\rm A}) \nn \\
{\cal J}^z:&&(\g^z,\overline\be_z),\quad (\be^z,\overline\g_z), \quad
(M^z,A_z)~, \label{epsval}
\end{eqnarray}
where $A=\lam^\pm,\xi,\ve,X\pm$ and $z=\nu_\pm,\Lam_\pm,X\pm$,
respectively, label the bosonic and fermionic constraints given
in (\ref{sVconst}). The classical BRST charge can then be constructed
directly from (\ref{sVconst}) and (\ref{csValg})
without recourse to gauge-fixing as
\begin{eqnarray}
Q&=&\int d\s \biggl[~{\cal C}^{\rm A}\vp_{\rm A}+\g^z{\cal J}_z
+{\cal P}^{\rm A}B_{\rm A}+\be^zA_z \nonumber \\
&&+{\cal C}^+(\overline{\cal P}_+{\cal C}^{+\prime}
+\overline\be_+\g^{+\prime})
+\g^+\biggl(2i\overline{\cal P}_+\g^+
-{1\over2}\overline\be_+{\cal C}^{+\prime}\biggr) \nonumber \\
&&-{\cal C}^-(\overline{\cal P}_-{\cal C}^{-\prime}
+\overline\be_-\g^{-\prime})
+\g^-\biggl(2i\overline{\cal P}_-\g^-
+{1\over2}\overline\be_-{\cal C}^{-\prime}\biggr)~\biggr]~,
\label{QB}
\end{eqnarray}
where ${\rm A}$ and $z$ run over all constraint labels. The
ghost trilinear terms are determined so that $Q$ satisfies
the nilpotency condition under the super-Piosson brackets as
\begin{equation}
\{Q~,~Q\}=0~. \label{nil}
\end{equation}
The dynamics are controlled by the BRST invariant total
hamiltonian $H_T$, which consists of the canonical hamiltonian
and the gauge-fixing term. In the present case the canonical part
vanishes identically, and $H_T$ is given by
\begin{equation}
H_T=\{Q,\Psi\}, \label{tH}
\end{equation}
where $\Psi$ is a gauge fermion. We shall use the standard form of
$\Psi$ given by
\begin{equation}
\V=\int d\s[~\overline{\cal C}_{\rm A}\chi^{\rm A}
+\overline\g^z\chi_z+\overline{\cal P}_{\rm A}N^{\rm A}
+\overline\be^zM_z~]~,
\label{gF}
\end{equation}
where $\chi^{\rm A}$ and $\chi_z$ denote the gauge conditions
imposed on dynamical variables. The BRST invariance of $H_T$ can be
stated as
\begin{equation}
\{Q, H_T\}=0~,\label{qht}
\end{equation}
which is automatically satisfied by (\ref{nil}).

So far our arguments are restricted within classical theory. In
quantum theory, the operators $Q$ and $H_T$ must be
suitably regularized, and the quantum version of the BRST invariance
(\ref{nil}) and (\ref{qht}) may fail to be valid due to anomalies.
Assuming that the anomalous commutators
\footnote{A careful analysis on the equal-time commutators appeared in the BRST
algebra is given in \cite{kubo}.} can be expanded in $\hbar$ as
\begin{eqnarray}
&& [Q~,~ Q] \equiv i\hbar^2\Omega+{\rm O}(\hbar^3)~, \nn \\
&& [ Q~ ,~ H_T] \equiv {i\over2}\hbar^2\Gamma+{\rm O}(\hbar^3), \label{anom}
\end{eqnarray}
we obtain the algebraic consistency condition for $\W$ in the lowest
order of $\hbar$ as
\begin{equation}
\de \Omega =0~,\qquad
\label{cstome}
\end{equation}
where $\de$ is the classical BRST transformation given by the
super-Poisson bracket $\de F = - \{Q~,~F \}$ for any $F$.
On the other hand, $\G$ can be related with $\W$ by
\begin{equation}
\Gamma = \{\Omega~ ,~ \Psi\}~.
\label{dirgam}
\end{equation}
Hence it is sufficient to cancel the anomaly in $Q^2$ to retain
BRST invariance.

The consistency condition (\ref{cstome}) has been solved in the EPS
\cite{fikkt}, and the nontrivial (matter independent) solution is found to be
\begin{equation}
\W=K~\int d\s\biggl[({\cal C}^{+}\dl_\s^3{\cal C}^{+}
-8i\g^{+}\dl_\s^2\g^{+})-({\cal C}^{-}\dl_\s^3{\cal C}^{-}
+8i\g^{-}\dl_\s^2\g^{-})\biggr]~.
\label{solution}
\end{equation}
The cohomology class to which this solution belongs is uniquely
fixed and independent of the choices of regularizations and
gauge-fixing.

The overall coefficient $K$, however, remains undetermined in
this algebraic method. To fix $K$, we must define operator products
by some ordering prescription, and then examine the nilpotency of
the BRST charge. For the purpose to define operator ordering, we
decompose operators into two parts by
\begin{eqnarray}
&&A^{(\pm)}(\s)=\int d\s\de^{(\mp)}(\s-\s')A(\s')
\quad{\rm for~~} A=P+X', \psi_+, \CC^+, \overline\CP_+, \nonumber \\
&&A^{(\pm)}(\s)=\int d\s\de^{(\pm)}(\s-\s')A(\s')
\quad{\rm for~~} A=P-X', \psi_-, \CC^-, \overline\CP_-, \label{opord}
\end{eqnarray}
with $\displaystyle{\de^{(\pm)}(\s)=\frac{1}{2\p}\frac{\pm i}{\s\pm\e}}$.
The $A^{(+)}$ and $A^{(-)}$ thus defined reduce, respectively, to
positive- and negative-frequency part in the superorthonormal gauge.
By putting $A^{(+)}$'s to the right of $A^{(-)}$'s, we define operator
ordering. We thus obtain
\begin{equation}
K=\frac{10-D}{16\p} \label{Kvalue}
\end{equation}
for the anomaly coefficient. This coincides with the result of
superconformal gauge-fixing \cite{imnu}.
If we change the ordering prescription,
the value of $K$ also changes. Which ordering should be employed depends
on the gauge chosen. Because of this, the operator ordering introduced
above should be understood as temporal. We will come back to this point
later.

The $Q^2$ anomaly is a direct consequence of the super-Virasoro anomaly of
the generalized super-Virasoro constraints defined by
\begin{eqnarray}
\F_\pm&\equiv&\vp^X_\pm\pm2\overline{\cal P}_\pm{\cal C}^{\pm\prime}
\pm\overline{\cal P}_\pm'{\cal C}^\pm
\pm{3\over2}\overline\be_\pm\g^{\pm\prime}
\pm{1\over2}\overline\be_\pm'\g^\pm \nonumber \\
I_\pm&\equiv&{\cal J}^X_\pm\mp{3\over2}\overline\be_\pm{\cal C}^{\pm\prime}
\mp\overline\be_\pm'{\cal C}^{\pm\prime}
+4i\overline{\cal P}_\pm\g^\pm ~,
\label{gsVcon}
\end{eqnarray}
where operator ordering is implicitly assumed. Suppressing $\hbar$
henceforth, we can easily show that these satisfy
\begin{eqnarray}
&&[\F_\pm(\s),\F_\pm(\s')]=\pm i(\F(\s)+\F(\s'))\dl_\s \de(\s-\s')
\pm iK\dl_\s^3\de(\s-\s')~, \nonumber \\
&&[I_\pm(\s),\F_\pm(\s')]
=\pm i{3\over2}I_\pm(\s)\dl_\s\de(\s-\s')
\pm iI'_\pm(\s)\cdot\de(\s-\s'), \nonumber \\
&&[I_\pm(\s),I_\pm(\s')]=4\F_\pm(\s)\de(\s-\s')
+8K\dl_\s^2\de(\s-\s'),\nonumber \\
&&{\rm all~other~super\hbox{-}commutators~vanish.} \label{qsValg}
\end{eqnarray}
One finds that due to the appearance of the anomalous Schwinger term, the
super-Virasoro constraints become second class ones.

The result (\ref{solution}) or (\ref{qsValg}) with $K$ given by
(\ref{Kvalue}) does not rely on $\hbar$-expansion and exact as far as
the $Q^2$ anomaly is concerned. Strictly speaking, the BRST anomaly
must be canceled in order for the higher order corrections to be
meaningful. The critical string with $D=10$ is such a case, where the
super-Liouville mode of 2D SUGRA multiplet is decoupled from the
theory. In noncritical strings, however, we must take account of the
super-Liouville mode as a dynamical variable, which also contributes
to the BRST anomaly.

Instead of taking account of the super-Liouville mode which is expected
to become dynamical in quantum theory, we modify the theory to recover
all the classical local symmetries violated by anomalies. This can
be carried out, without affecting the physical contents of the original
theory, by introducing extra degrees of freedom which can be formally
gauged away by the recovered local symmetry.

Following the general idea of BF \cite{bf}, we introduce a canonical
pair of bosonic fields $(\theta, \p_\theta)$ and a Majorana
field $\z_\pm$, which we will refer to as BF fields henceforth.
They are assumed to satisfy the same type of canonical
supercommutation relations as string variables. We then modify
the constraints (\ref{gsVcon}) by adding to $\Phi_\pm$ and $I_\pm$
an appropriate terms containing BF fields to cancel the
super-Virasoro anomaly in (\ref{qsValg}). Let us denote the modified
super-Virasoro operators by $\tilde\Phi_\pm$ and $\tilde I_\pm$, then they
are given by
\begin{eqnarray}
\tilde{\Phi}_{\pm}&\equiv& \Phi_{\pm}+\frac{1}{4}\Theta_\pm^2
-\g\Theta_{\pm}^\prime\pm\frac{i}{2}\z_{\pm}\z_{\pm}^{\prime}
+\g^2\mu^2e^\frac{\th}{\g}\mp\frac{i\mu}{2}e^\frac{\th}{2\g}
\z_\mp\z_\pm ~, \nn \\
\tilde{I}_{\pm}&\equiv& I_{\pm}\pm\z_{\pm}\C_{\pm}\mp4\g\z_{\pm}'
\mp2\g\mu e^\frac{\th}{2\g}\z_\mp ~,
\label{modconst}
\end{eqnarray}
where we have defined $\Theta_\pm\equiv \theta'\pm\p_\theta$. The $\g$
is a free parameter to be fixed to cancel the super-Virasoro anomaly.
In fact, the super-Virasoro constraints of the BF sector satisfy the
super-Virasoro algebra with anomaly coefficient
\begin{equation}
\k\equiv 2\g^2 \label{kappa}
\end{equation}
under classical super-Poisson brackets. Since the BF fields also
contribute to the super-Virasoro anomaly as a single superconformal
matter multiplet in quantum theory, the BRST anomaly can be canceled
if $\k$ satisfies
\begin{equation}
\frac{D-10+1}{16\p}+\k=0~. \label{mcoeff}
\end{equation}
In (\ref{modconst}), the terms containing the mass parameter $\mu^2$ is
not necessary for the purpose to cancel the BRST anomaly but they turn
out to be related with the cosmological terms in the covariant effective
action as we shall see in the next section. In quantum theory,
the exponential operators $\displaystyle{\exp\frac{\th}{2\g}}$
will be modified by gravitational dressing effect \cite{ct,d,dk}. The full
quantum mechanical treatment of them will be discussed in Section 4.

As we mentioned above, different choice of operator ordering gives rise to
different anomaly coefficient. For the superconformal gauge-fixing,
the ordering prescription introduced above can be used and (\ref{mcoeff})
turns out to be correct. In the light-cone gauge, we will adopt
different ordering and arrive at different conditions for $\k$ as we
will see in Section 5.

\section{Effective action and geometrization}
\setcounter{equation}{0}

In the previous section we have introduced the BF fields, and modified
the super-Virasoro constraints so as to cancel the BRST anomaly.
In this section we will apply the BFV algorithm to the gauge
symmetrized system and investigate the BRST invariant effective
action.

Let us denote the BRST charge modified by the BF fields by $\tilde Q$,
then it generates the BRST transformation of any variable $F$ by
\begin{equation}
\de F=i[\tilde Q,F]~. \label{tilBRST}
\end{equation}
We thus obtain the following set of BRST transformations in the EPS;
\begin{eqnarray}
&& \de X={1\over2}\{({\cal C}^+-{\cal C}^-)X'+({\cal C}^++{\cal C}^-)P\}
+\g^+\psi_++\g^-\psi_-,\nn \\
&& \de P=\biggl({1\over2}\{({\cal C}^+-{\cal C}^-)P+({\cal C}^+
+{\cal C}^-)X'\}+\g^+\psi_+-\g^-\psi_-\biggr)',\nn \\
&& \de\psi_\pm=\pm{1\over2}{\cal C}^{\pm\prime}\psi_\pm\pm{\cal C}^\pm\psi_\pm'
+i\g^\pm(P\pm X'),\nn \\
&& \delta \th ={1\over 2}(~{\cal C}^{+}\Theta_+
-{\cal C}^-{\Theta_-}~)-\g({\cal C}^{+} -{\cal C}^{-})'
+\g^+\z_++\g^-\z_- ~,\nonumber\\
&& \delta \pi_{\th} =\biggl(~\frac{1}{2}({\cal C}^{+}
\Theta_+-{\cal C}^{-}\Theta_-)-\g({\cal C}^{+} +{\cal C}^{-})'
+\g^+\z_+-\g^-\z_-\biggr)'~,  \nonumber\\
&& \de \z_{\pm}=\pm{\cal C}^{\pm}\z_{\pm}^{\prime}\pm
\frac{1}{2}{\cal C}^{\pm \prime}\z_{\pm}\pm i\g^{\pm}\Theta_{\pm}
\mp4i{\g}\g^{\pm \prime}, \nn \\
&& \de\lam^\pm={\cal C}_\lam^\pm, \qquad \de\x={\cal C}^\x,
\qquad \de\ve={\cal C}^\ve, \nn \\
&& \de{\cal C}_\lam^\pm=0, \qquad
\de{\cal C}^\x=0, \qquad \de{\cal C}^{\ve}=0, \nn \\
&& \de\n_\pm=-\g^\n_\pm, \qquad \de\Lam_\pm=-\g^\Lam_\pm, \qquad
\de\g^\n_\pm=0, \qquad \de\g^\Lam_\pm=0, \label{brsttr}\\
&& \de N^A={\cal P}^A, \qquad \de M_z=-\be_z, \qquad
\de{\cal P}^A=0, \qquad \de\be^z=0~, \nn \\
&& \de\overline{\cal C}_A=-B_A, \qquad
\de\overline\g^z=-A^z, \qquad \de B_A=0, \qquad \de A^z=0, \nn \\
&& \de\overline{\cal P}_{\hat A}=-\vp_{\hat A}, \qquad
\de\overline\be^{\hat z}=-{\cal J}^{\hat z} \qquad
\de\overline{\cal P}_{\pm}=-\tilde\Phi_{\pm}, \qquad
\de\overline\be_{\pm}=-\tilde{I}_{\pm}, \nn \\
&& \de{\cal C}^\pm=\pm{\cal C}^\pm{\cal C}^{\pm\prime}-2i(\g^\pm)^2, \qquad
\de\g^\pm=\mp{1\over2}{\cal C}^{\pm\prime}\g^\pm
\pm{\cal C}^\pm\g^{\pm\prime} ~, \nn
\end{eqnarray}
where the constraint labels $\hat A$ and $\hat z$ run only through primary
constraints, while $A$ and $z$ are taken all over the bosonic and fermionic
constraints.

The change in the BRST charge is reflected on the dynamics through
the total hamiltonian
\begin{eqnarray}
\tilde{H}_T&=&\frac{1}{i}[\tilde{Q}~,~\Psi]~.
\label{newhami}
\end{eqnarray}
This is BRST invariant if the $\tilde Q^2$ anomaly is absent.
To construct the effective action we choose the standard form of
gauge fermion (\ref{gF}) shifted by $\Psi\rightarrow\int d\s[\overline\CC_A\dot
N^A+\overline \g^z\dot M_z$.
This just cancels the Legendre terms $\int d^2x[\overline\CC_A\dot\CP^A
+\overline\g^z\dot\be_z+B_A\dot N^A+A^z\dot M_z]~$ in constructing the
effective action. The BRST invariant effective action can be obtained  as
\begin{equation}
 S_{\rm eff}=S_0+S_{\rm gh}+S_{\rm gf},
 \label{gfa}
\end{equation}
where
\bn
S_0&=& \int d^2x[~P\dot X+{i\over2}(\psi_+\dot\psi_+
+\psi_-\dot\psi_-)+\pi_\th\dot\th+{i\over2}(\z_+\dot\z_+
+\z_-\dot\z_-) \nn \\
&& +\p^\lam_+\dot\lam^++\p^\lam_-\dot\lam^-
+\p_\x\dot\x+\p_\ve\dot\ve+\p^+_\n\dot\n_++\p^-_\n\dot\n_-
+\p^-_\Lam\dot\Lam_++\p^-_\Lam\dot\Lam_- \nn \\
&& -N^{\hat A}\vp_{\hat A}-N^+\tilde\vp_+-N^-\tilde\vp_-
+M_{\hat{z}}{\cal J}^{\hat{z}}+M^+\tilde\CJ_++M^-\tilde\CJ_-
]~, \label{S0} \\
S_{\rm gh}&=&\int d^2x[\overline{\cal P}_A\dot{\cal C}^A
+\overline\be^z\dot\g_z-\overline{\cal C}_A\de\chi^A
+\overline\g^z\de\chi_z-\overline{\cal P}_A{\cal P}^A
-\overline\be^z\be_z \nn \\
& &-N_+(2\overline{\cal P}_+{\cal C}^{+\prime}
+\overline{\cal P}'_+{\cal C}^++{3\over2}\overline\be_+\g^{+\prime}
+{1\over2}\overline\be'_+\g^+) \nn \\
& &+N_-(2\overline{\cal P}_-{\cal C}^{-\prime}
+\overline{\cal P}'_-{\cal C}^-+{3\over2}\overline\be_-\g^{-\prime}
+{1\over2}\overline\be'_-\g^-) \label{Seff}\\
& &-M^+({3\over2}\overline\be_+{\cal C}^{+\prime}+\overline\be'_+{\cal C}^+
-4i\overline{\cal P}_+\g^+) \nn \\
& &+M^-({3\over2}\overline\be_-{\cal C}^{-\prime}+\overline\be'_-{\cal C}^-
+4i\overline{\cal P}_-\g^-)]~, \nn \\
S_{\rm gf}&=&\int d^2x[-B_A\chi^A-A^z\chi_z]~. \label{Sgf}
\en
In (\ref{S0}) $\tilde\vp_\pm$ and $\tilde\CJ_\pm$, respectively, are obtained
from the super-Virasoro constraints $\tilde\Phi_\pm$ and $\tilde\CJ_\pm$
given by (\ref{modconst}) with all the ghost contributions removed.
The $\chi^A$ and $\chi_z$ stand for the gauge-fixing conditions.
We have used the BRST transformations (\ref{brsttr}) in deriving these
actions.

For a wide class of gauge conditions the effective action provides us with
a starting point to analyze the quantum theory of fermionic string as
2D SUGRA. It is, however, written in terms of the EPS variables,
which have no direct geometrical interpretation unless we explicitly
specify the gauge conditions. In order to see the physical
significance of the effective action, we relate the EPS variables with
the configuration space variables to geometrize the effective action.
The geometrization of the EPS variables has been developed in
ref. \cite{fikm,fikkt}. No essential change will appear in relating the ghost
variables in the EPS with those of configuration space in the
presence of the BF variables.

Since the auxiliary fields $N^\pm$ and
$M^\pm$, respectively, play the roles of $\lam^\pm$ and $\pm i\nu_\mp$
in (\ref{S0}), we identify them by imposing the gauge conditions
\begin{equation}
\chi^\pm_\lam=\lam^\pm-N^\pm, \qquad \chi_\pm^\nu=\nu_\pm\mp iM^\mp~.
\label{gcond}
\end{equation}
As for the rest of the gauge conditions, we assume that they and
their BRST transforms are independent of $P$, $\psi_\pm$, $\p_\th$,
$\z_\pm$, $\p_{\hat A}$, $\p^{\hat z}$, $\overline\CP_A$ and
$\overline\be^z$. They are otherwise arbitrary. This is suffice for
the purpose of geometrization.

Due to the assumptions on the gauge conditions, we can derive equations
of motion for $P$, $\psi_\pm$, $\p_\th$, $\z_\pm$, $\p_{\hat A}$,
$\p^{\hat z}$, $\overline\CP_A$ and $\overline\be^z$ by taking the
variation of (\ref{gfa}) with respect to these variables as
\begin{eqnarray}
&& \dot X-{1\over2}\{(N^++N^-)P+(N^+-N^-)X'\}+M^+\psi_++M^-\psi_-=0, \nn \\
&&\dot{\th} -{1\over2}\{(N^++N^-)\p_{\th}+(N^+-N^-)\th'~\}
-\g (N^+-N^-)^\prime+M^+\z_++M^-\z_-=0, \nn \\
&& \dot\psi_\pm\mp N^\pm\psi'_\pm\mp{1\over2}N^\pm{}'\psi_\pm
+iM^\pm(P\pm X')=0, \nn \\
&& \dot\z_\pm\mp N^\pm\z'_\pm\mp{1\over2}N^\pm{}'\z_\pm
+iM^\pm(\p_{\th}\pm \th')\mp4i\g M^\pm{}'=0, \nn \\
&& \dot\lam^\pm=N^\pm_\lam, \qquad \dot\x=N^\x, \qquad \dot\ve=N^\ve,
\qquad \dot\n_\pm=M^\n_\pm, \qquad \dot\Lam_\pm=M^\Lam_\pm, \label{eom}\\
&& \dot{\cal C}_\lam^\pm={\cal P}^\pm_\lam,\qquad
\dot{\cal C}^\x={\cal P}^\x,\qquad
\dot{\cal C}^\ve={\cal P}^\ve, \qquad
\dot \g^\n_\pm=\be^\n_\pm, \qquad \dot\g^\Lam_\pm=\be^\Lam_\pm, \nn \\
&& {\cal P}^\pm=\dot{\cal C}^\pm\pm{\cal C}^\pm N^\pm{}'
\mp{\cal C}^\pm{}'N^\pm-4i\g^\pm M^\pm, \nn\\
&& \be^\pm=\dot\g^\pm\pm{1\over2}\g^\pm N^\pm{}'\mp\g^\pm{}'N^\pm
\mp{\cal C}^\pm M^\pm{}'\pm{1\over2}{\cal C}^\pm{}'M^\pm~.
\end{eqnarray}
We require that if we use the equations of motion (\ref{eom}) to
eliminate $P$, $\CP^\pm$ and $\be^\pm$, the BRST transformations
(\ref{brsttr}) in the EPS reduce to those in the configuration space
given in Appendix. For instance, the covariant reparametrization ghosts
$C^\al$ and the supersymmetry ghosts $\w$ can be identified by
comparing the BRST transformations of $X$ in the EPS and
in the configuration space in (\ref{ncbrsttr}). They are given by
\begin{eqnarray}
&&C^0={{\cal C}^++{\cal C}^-\over N^++N^-}, \qquad
C^1={N^-{\cal C}^+-N^+{\cal C}^-\over N^++N^-}~, \nn \\
&&\w=\pmatrix{e^{\frac{\x}{2}+\ve}\w_-\cr e^{\frac{\x}{2}-\ve}\w_+}
\quad{\rm with} \quad
\w_\pm=\mp i\biggl(\g^\mp
+{{\cal C}^++{\cal C}^-\over N^++N^-}M^\mp\biggr)~.
\label{supergh}
\end{eqnarray}
The Weyl ghost $C_W$ and the local Lorentz ghost $C_L$ can be found
from the BRST transformation properties of $\xi$ and $\ve$ to be
\bn
C_W &=&{\cal C}^\x-V^+_{\cal C}+V^-_{\cal C}~, \nn \\
C_L &=& \CC^\ve-Z_\CC^++Z_\CC^-~,   \label{clcw}
\en
where $V_\CC^\pm$ and $Z_\CC^\pm$ are defined by
\begin{eqnarray}
V_{\cal C}^\pm&=&{1\over2}G_\pm{\cal C}^\pm\pm\Lam_\pm\g^\pm
+{\cal C}^{\pm\prime}~, \nn \\
Z_\CC^\pm&=&\pm\frac{1}{2}\CC^\pm L_\pm
+\frac{1}{2}\g^\pm\Lam_\pm\pm\frac{1}{2}\CC^\pm{}' \label{VCpm}
\end{eqnarray}
with
\begin{eqnarray}
G_\pm&=&{2\over N^++N^-}[\pm N^\x+N^\mp\x'\mp(N^+-N^-)'
\mp(\Lam_+M^++\Lam_-M^-)]~, \nn \\
L_\pm&=&{2\over N^++N^-}[\pm N^\ve+N^\mp\ve'\mp(N^++N^-)'
\mp(\Lam_+M^+-\Lam_-M^-)]~.
\end{eqnarray}
We finally obtain the super-Weyl ghost from the BRST transformations
of $\Lam_\pm$ as
\begin{equation}
\h_{\scriptscriptstyle W}
=\pmatrix{e^{\ve-\frac{\x}{2}}
\h_{{\scriptscriptstyle W}-} \cr
e^{-\ve-\frac{\x}{2}}\h_{{\scriptscriptstyle W}+}} \quad{\rm with} \quad
\h_{{\scriptscriptstyle W}\pm}=W_{\cal C}^\pm
+{{\cal C}^++{\cal C}^-\over N^++N^-}W_N^\pm~, \label{itaw}
\end{equation}
where $W_N^\pm$ and $W_{\cal C}^\pm$ are defined by
\begin{eqnarray}
W_N^\pm&=&M^\Lam_\pm\mp\Lam'_\pm N^\pm\mp{1\over2}\Lam_\pm N^\pm{}'
\pm i(G_\pm M^\pm+4M^\pm{}') ~,\nn\\
W_{\cal C}^\pm&=&\g^\Lam_\pm\mp\Lam'_\pm{\cal C}^\pm
\mp{1\over2}\Lam_\pm{\cal C}^\pm{}'\pm i(G_\pm\g^\pm+4\g^\pm{}')~.
\end{eqnarray}
The equations (\ref{supergh}), (\ref{clcw}) and (\ref{itaw}) fix
the ghost relations between the BFV basis and the covariant one.

Except for the BF fields, which we will shortly discuss, it is
straightforward to convert the BRST transformations of the
remaining variables in terms of configuration space variables.
Furthermore, the BRST transformations of the covariant ghost
variables can be easily obtained from (\ref{brsttr}) and
(\ref{eom}). We shall
introduce here auxiliary fields $F_X^\mu$, $F_G$, $F_L$
and $F_W$ for the supermultiplets corresponding to the string variables,
the gravity, the Liouville fields and the super-Weyl ghost,
respectively.
These are needed to close the algebra
off-shell.
The BRST transformations constructed from
(\ref{brsttr}) by using (\ref{eom}) and the ghost relations
given above coincide with those of Appendix up to the terms
containing auxiliary fields, being only on-shell nilpotent.
The original BRST
transformations (\ref{brsttr}) in the EPS satisfy the off-shell nilpotency
by construction. There, the number of bosonic fields
balances that of fermionic fields. Eliminating the momentum
variables, however, destroys the balance, and the off-shell
nilpotency is no longer satisfied.
Incidentally, the off-shell
nilpotency of these two BRST transformations is realized in a
quite different manner. In the configuration space, the covariance
is manifestly maintained and the nilpotency is consistent with the
covariance. On the other hand, the covariance is only manifest
on the mass-shell in the EPS
and the off-shell nilpotency is realized by sacrificing the
manifest covariance.
Therefore, when passing to the configuration space, we include auxiliary
fields to retain the off-shell nilpotency needed to investigate
the theory in arbitrary gauges.

With these preparation, we turn to the geometrization of the BF
fields. In terms of covariant ghost variables the BRST
transformations of the BF fields are given by
\begin{eqnarray}
\de \th&=&C^\al\partial_\al\th-i(\w_-\z_+-\w_+\z_-)
-\g(2C^{1\prime}+C^{0\prime}(\lam^+-\lam^-))~, \nn \\
\de\z_\pm&=&C^\al\partial_\al\z_\pm+
\frac{1}{2}(C^1{}'\pm\lam^\pm C^0{}')\z_\pm
-4\g C^0{}'\nu_\mp-4\g\w'_\mp \nn \\
&&\pm\frac{2\w_\mp}{\lam^++\lam^-}[\dot\th\pm\lam^\mp\th'
+\g(\lam^+-\lam^-)'+i(\nu_-\z_+-\nu_+\z_-)]~. \label{deth}
\end{eqnarray}
The presence of the terms proportional to $\g$ implies that the
$\th$ and $\z_\pm$  have no simple transformation properties as
scalar and spinor components of a scalar supermultiplet for the
string variables. Rather, the BF fields possess transformation
properties similar to those of $\xi$ and $\Lam_\pm$ as is easily
seen from (\ref{ncbrsttr}). In fact we can construct covariants by taking
the combinations
\bn
\f&=&\x - \frac{1}{\g}\th~,\nn \\
\h_{}&=&\pmatrix{e^{\ve-\frac{\x}{2}}\h_{-}\cr
e^{\ve-\frac{\x}{2}}\h_{+} \cr}
\quad{\rm with}\quad \h_\pm=\Lam_\pm-\frac{1}{\g}\z_\pm ~.
\label{Liouv}
\en
The covariant BRST transformations of these variables coincide
with those of the super-Liouville multiplet given in Appendix,
where we have included the auxiliary fields $F_L$ as mentioned
above. We see from this that $(\f,\h,F_L)$ can be regarded as
super-Liouville multiplet. They not only transform as a scalar
multiplet under the reparametrizations and supersymmetry but
also change under the super-Weyl transformations.

The master action (\ref{gfa}) contains many nonpropagating
degrees, i.e., $P$, $\p_\th$, $\CP^A$, $\overline\CP_A$, $\be_z$,
$\overline\be^z$, $N^A$ and $M_z$, which can be eliminated by
virtue of the equations of motion (\ref{eom}). After eliminating
these variables from the master action and rewriting it in terms
of the covariant variables defined above, we arrive at the
gauge-fixed covariant action
\begin{equation}
S_{\rm eff}=S_X+S_\f+S_g+S_{\rm cosm}+S_{\rm aux}
+S_{\rm gh}+S_{\rm gf} \label{gfca}
\end{equation}
with
\begin{eqnarray}
S_\f&=&\frac{\k}{2}S^{\f,\h}_L~, \nn \\
S_g&=&-\frac{\k}{2}\Bigl[S^{\x,\Lam}_L
+\int d^2x eg^{00}\Bigl\{\Bigl(\frac{g_{01}}{g_{11}}\Bigr)'\Bigr\}^2\Bigr]~,
\nn \\
S_{\rm cosm}&=&-\frac{\k}{2}\mu\int d^2x ee^{-\frac{\f}{2}}\Bigl[
-2(F_G-F_L)+\frac{1}{2}\overline\h\h-2i
\overline\chi_\al\r^\al\h
+4\e^{\al\be}\overline\chi_\al\r_5\chi_\be \Bigr]~, \nn \\
S_{\rm aux}&=&\frac{1}{2}\int d^2x eF_X^2+\frac{\k}{4}\int d^2x e(F_G-F_L)^2~,
\nn \\
S_{\rm gh}&=& \int d^2x[-\overline\CC_+\de\chi^+_\vp
-\overline\CC_-\de\chi_\vp^--\overline\CC_\x\de\chi^\x
-\overline\CC_\ve\de\chi^\ve -\overline\CC_f\de \chi^f \nn \\
&&+\overline\g_+\de\chi_\CJ^++\overline\g_-\de\chi_\CJ^-
+\overline\g^+_\Lam\de\chi_+^\Lam+\overline\g^-_\Lam\de\chi_-^\Lam]~, \nn \\
S_{\rm gf}&=&\int d^2x[ -B_+\chi_\vp^+-B_-\chi_\vp^--B_\x\chi^\x
-B_\ve\chi^\ve -B_f\chi^f \nn \\
&& -A_+\chi^+_\CJ-A_-\chi^-_\CJ-A^+_\Lam\chi_+^\Lam-A^-_\Lam\chi_-^\Lam]
\label{SLetc}
\end{eqnarray}
where $S^{\f,\h}_L$ is supersymmetric Liouville action given by
\bn
S^{\f,\h}_L&=&\int d^2x~e\biggl[~-\frac{1}{2}
\bigl(g^{\al\be}\dl_\al \phi\dl_\be \phi
-i\overline\h\r^\al\nabla_\al\h\bigr)
-\overline\chi_\al\r^\be\r^\al\h\dl_\be \phi
-\frac{1}{4}\overline\h\h\overline\chi_\al\r^\be\r^\al
\chi_\be \nonumber\\
&&+R\phi
+4i\e^{\al\be}\overline\chi_\al\r_5\r^\g\chi_\g\partial_\be\phi
+4\e^{\al\be}\overline\chi_\al\r_5\nabla_\be\h~\biggr]~ \label{sLact}
\en
with $R$ being the scalar curvature of the metric $g_{\al\be}$.
The $S^{\x,\Lam}_L$ can be obtained from (\ref{sLact}) by the replacement
$\f\rightarrow \x$ and $\h_\pm \rightarrow \Lam_\pm$. This is a local
functional of 2D SUGRA fields and can be considered
a super-Liouville action with
\begin{equation}
\xi=\ln g_{11}~, \qquad
\Lam=\pmatrix{4e^{2\ve-\xi}\chi_{1+} \cr
4e^{-2\ve-\xi}\chi_{1-}}  \label{sLiouvmod}
\end{equation}
as the super-Liouville fields.

In (\ref{gfca}) we have also included the  auxiliary fields to retain the
BRST invariance in the configuration space given in Appendix.
In this connection, we notice here appearance of
a new local symmetry associated with $F_{G}(x) \rightarrow
 F_{G}(x) + \lam(x), ~F_{L}(x) \rightarrow F_{L}(x) + \lam(x)$
by an arbitrary function $\lam(x)$. To gauge-fix the local symmetry, we have
supplemented a new gauge condition $\chi^f$ and an anti-ghost
$\overline\CC_f$ to (\ref{gfca}). The $F_W$ required by the off-sell
nilpotency of the BRST transformation in the super-Weyl ghost sector
can be considered as the ghost corresponding to this symmetry,
as can be seen from the BRST transformations of $F_G$ and $F_L$
given in (\ref{cbrsttr}).

It is very interesting to see the properties of (\ref{gfca}) under the
classical local symmetries of (\ref{str}).
As was discussed in ref. \cite{fikkt},
the $S_g$ is not invariant under the classical symmetries (except for the
local Lorentz invariance) and produces both super-Virasoro and super-Weyl
anomalies even in the classical theory. The quantization of the string
variables breaks the reparametrization invariance and the local
supersymmetry, and leads to the super-Virasoro anomaly. This anomaly
together with the contributions from other sectors including the
supergravity itself is canceled by the classical super-Virasoro
anomaly produced by $S_g$. The super-Weyl anomaly of $S_g$
is canceled by that of $S_\f$. This is rather obvious since the
combination $S_\f+S_g$ depends on $\x$, $\f$, $\Lam_\pm$ and $\h_\pm$
only through the BF fields which are invariant under the super-Weyl
transformations. Furthermore, the cosmological term can be obtained
by eliminating
the auxiliary by the equation of motion $F_G-F_L=-2\mu e^{-\frac{\f}{2}}$
as
\begin{equation}
-\frac{\k}{2}\int d^2x e\Bigl[
2\mu^2 e^{-\f}+\frac{\mu}{2}\overline\h\h e^{-\frac{\f}{2}}-2i\mu
\overline\chi_\al\r^\al\h e^{-\frac{\f}{2}}
+4\mu\e^{\al\be}\overline\chi_\al\r_5\chi_\be e^{-\frac{\f}{2}}\Bigr]~.
\label{Scosm}
\end{equation}
It is easy to show that this is invariant under the super-Weyl
transformations.

We now come back to the invariance of the effective action (\ref{gfca})
under the BRST transformations given in Appendix. From the very
construction of our symmetrization procedure, it is invariant under
the transformations without the auxiliary fields. When they
are included, the BRST transformations of Majorana fields are modified by
\begin{equation}
\de_F\psi=\w F_X~,\qquad
\de_F \chi_\al=-\frac{i}{4}\rho_{\al} \w F_G~, \qquad
\de_F \h=\w F_L~. \label{deF}
\end{equation}
It is well-known that the noninvariance of $S_X$ is canceled
by the variation of the $F_X^2$ term in $S_{\rm aux}$. Noting that
the modifications due to $F_G$ can be regarded as a fermionic
transformation, we easily find
\begin{eqnarray}
&&\de_FS_\f=\frac{\k}{2}\int d^2x e[i\overline\w\r^\al\{
\nabla_\al\h+i\r^\be(\partial_\be\f-\overline\chi_\be\h)\chi_\al\}(F_L-F_G)
-4\e^{\al\be}\overline\w\r_5\nabla_\al\chi_\be F_L]~, \nn \\
&&\de_FS_g=2\k\int d^2x e\e^{\al\be}
\overline\w\r_5\nabla_\al\chi_\be F_G~. \label{defsg}
\end{eqnarray}
These can be shown to cancel exactly the BRST transformation of
$S_{\rm aux}$. The fact that only the difference $F_G-F_L$ appears in
the action is related with the extra symmetry mentioned above.
The action (\ref{str}) supplemented by the auxiliary field $F_X$ by
itself is classically invariant under the full BRST transformations
containing $F_G$. The auxiliary filed $F_G$, however, enters into the
action only through quantum effects, i.e., the super-Weyl anomaly \cite{hu}.

We emphasize that though the effective action (\ref{gfca}) has been
derived by assuming the restricted class of gauge conditions stated
below (\ref{gcond}), it is considered to be valid for arbitrary gauge
conditions. This is because the effective action is invariant under
the BRST transformations maintaining the off-shell nilpotency.
If we restrict ourselves to the gauge conditions satisfying the
assumption stated below (\ref{gcond}), we need not introduce the
auxiliary fields. In such a case, the BRST transformations are only
on-shell nilpotent, but the equations of motion can be consistent
with the BRST symmetry. Actually this happens in the superconformal gauge.
The restriction on the gauge conditions, however,
is too strong to allow interesting gauges such as the light-cone gauge.
Let us discuss now these specific gauge fixings.

\section{Superconformal gauge-fixing}
\setcounter{equation}{0}

In the previous section we have formulated the BRST invariant
effective action. So far our argument does not rely on particular
gauge conditions. The effective action (\ref{gfa}) or its covariantized
version (\ref{gfca}) can be applicable to any gauge fixing. It is, however,
instructive to describe explicit calculations and illustrate
some issues which arise in quantizing the effective action. In this
section we will argue superconformal gauge-fixing.

Since the effective action (\ref{gfca}) possesses all the classical
local symmetries, we can fix the zweibeins and the gravitinos to
arbitrary background fields. The gauge-fixed action can be easily
constructed in this general case. In particular we can choose the
background to be flat minkowskian superspace. Here we shall discuss
this simplest case, i.e., $e_\al{}^a=\de_\al^a$, $\chi_\al=0$.
In terms of (\ref{zweibein}) the flat superorthonormal gauge can be
implemented by the following set of gauge conditions
\bn
&&\chi^{\pm}_\vp=N^{\pm}-1, \qquad \chi^{\ve}=\ve, \qquad
\chi^{\x}=\x \nn \\
&&\chi^{\pm}_\CJ=\mp iM^{\pm}, \qquad \chi_{\pm}^{\Lam}
=\Lam_{\pm}~, \qquad\chi^f=f_G~, \label{scgcd}
\en
where $\chi^{\pm}_\vp$ and $\chi^{\pm}_\CJ$ denote the gauge
conditions for the super-Virasoro constraints. The last gauge condition
in (\ref{scgcd}) is to fix the local symmetry associated with the
presence of the auxiliary fields discussed in the previous section.
In this gauge the Lorentz ghost, the super-Weyl ghosts can be related
with the superreparametrization ghosts as
\begin{equation}
C_L=-\frac{1}{2}\e^{\al\be}\dl_\al C_\be ~, \qquad
C_W=-\dl_\al C^\al ~, \qquad
F_W=0 ~, \qquad
\h_{\scriptscriptstyle W}=-2i\r^\al\dl_\al\w ~,
\label{CLWghosts}
\end{equation}
and their anti-ghosts vanish. Integrating out the multipliers $B_A$,
$A^z$, the auxiliary fields $F_{X,L}$, and eliminating nonpropagating
variables by the equations of motion, we can reduce the effective action
(\ref{gfca}) to the following expression\footnote{The light-cone coordinates
are denoted by $x^\pm=x^0\pm x^1$. Correspondingly, the flat metric is
given by $\h_{++}=\h_{--}=0$, $\h_{+-}=\h_{-+}=-\frac{1}{2}$. We
exceptionally define the derivatives by $\dl_\pm=\dl_0\pm\dl_1$, hence
$\dl_\pm x^\pm=2$.}
\begin{eqnarray}
S_{\rm eff}&=&\int d^2x\Bigl[-\frac{1}{2}\dl_+X\dl_-X
+\frac{i}{2}\psi_+\dl_-\psi_++\frac{i}{2}\psi_-\dl_+\psi_-\Bigr] \nn \\
&& +\frac{\k}{2}\int d^2x\Bigl[-\frac{1}{2}\dl_+\f\dl_-\f
+\frac{i}{2}\h_+\dl_-\h_++\frac{i}{2}\h_-\dl_+\h_-
-2\m^2e^{-\f}+i\m\h_-\h_+e^{-\frac{\f}{2}}\Bigr] \nn \\
&& +\int d^2x [ -b_{++}\dl_-C^+-b_{--}\dl_+C^-
+\be_{++}\dl_-\w_-+\be_{--}\dl_+\w_+] ~, \label{gfascg}
\end{eqnarray}
where the anti-ghosts $\overline\CC_\pm$ and $\overline\g_\pm$ are,
respectively, denoted by $b_{\pm\pm}$ and $\be_{\pm\pm}$. Then the
string variables and the ghosts become free fields, and the BF fields
satisfy the supersymmetric Liouville equations, i.e.,
\begin{eqnarray}
&& \dl_+\dl_-X=\dl_\pm\psi_\mp=0 ~, \qquad
\dl_\mp C^\pm=\dl_\mp\w_\mp=\dl_\mp b_{\pm\pm}=\dl_\mp\be_{\pm\pm}=0 ~,
\nn \\
&& \dl_+\dl_-\f-2\m^2e^{-\f}+\frac{i\m}{2}\h_-\h_+e^{-\frac{\f}{2}}=0 ~,
\qquad \dl_\pm\h_\mp\pm\h_\pm e^{-\frac{\f}{2}}=0 ~.
\label{sceom}
\end{eqnarray}

The gauge-fixed action (\ref{gfascg}) is invariant under the
BRST transformations
\begin{eqnarray}
\de X&=& \frac{1}{2}C^+\dl_+X+\frac{1}{2}C^-\dl_-X
-i(\w_-\psi_+-\w_+\psi_-) ~, \nn \\
\de \psi_\pm&=&\frac{1}{2}C^\pm\dl_\pm\psi_\pm
+\frac{1}{4}\dl_\pm C^\pm\psi_\pm\pm\w_\mp\dl_\pm X ~, \nn \\
\de\f&=&-\frac{1}{2}\dl_+C^+-\frac{1}{2}\dl_-C^-
+\frac{1}{2}C^+\dl_+\f+\frac{1}{2}C^-\dl_-\f
-i(\w_-\h_+-\w_+\h_-) ~, \nn \\
\de \h_\pm&=&\mp2\dl_\pm\w_\mp+\frac{1}{2}C^\pm\dl_\pm\h_\pm
+\frac{1}{4}\dl_\pm C^\pm\h_\pm\pm\w_\mp\dl_\pm\f
\pm\frac{\m}{2}C^\mp\h_\mp e^{-\frac{\f}{2}}
+2\m\w_\pm e^{-\frac{\f}{2}} ~, \nn \\
\de C^\pm&=&\frac{1}{2}C^\pm\dl_\pm C^\pm+2i\w_\mp^2 ~, \nn \\
\de\w_\pm&=&\frac{1}{2}C^\mp\dl_\mp \w_\pm
-\frac{1}{4}\dl_\mp C^\mp\w_\pm ~, \nn \\
\de b_{\pm\pm}&=&T^X_{\pm\pm}+T^L_{\pm\pm}+T^{gh(2)}_{\pm\pm}
+T^{gh(3/2)}_{\pm\pm} ~, \nn \\
\de \be_{\pm\pm}&=&\pm i(J^X_{\pm\pm}+J^L_{\pm\pm}+J^{gh}_{\pm\pm}) ~,
\label{scbrsttr}
\end{eqnarray}
where $T^{X,L,gh(2,3/2)}_{\pm\pm}$ and $J^{X,L,gh}_{\pm\pm}$ are the
components of stress tensors and the supercurrents of the string,
Liouville and ghost sectors. They are given by
\begin{eqnarray}
T^X_{\pm\pm}&\equiv&\vp^X_\pm
=\frac{1}{4}(\dl_\pm X)^2+\frac{i}{4}\psi_\pm\dl_\pm\psi_\pm ~, \nn \\
T^L_{\pm\pm}&\equiv&\frac{\k}{2}\Bigl[\frac{1}{4}(\f'\pm\frac{2}{\k}\p_\f)^2
-(\f'\pm\frac{2}{\k}\p_\f)'\pm\frac{i}{2}\h_\pm\h_\pm'+\m^2e^{-\f}
\mp\frac{i\m}{2}e^{-\frac{\f}{2}}\h_\mp\h_\pm\Bigr] \nn \\
&=&\frac{\k}{2}\Bigl[\frac{1}{4}(\dl_\pm\f)^2
+\frac{1}{2}\dl_\pm^2\f+\frac{i}{4}\h_\pm\dl_\pm\h_\pm\Bigr] ~, \nn \\
T^{gh(2)}_{\pm\pm}&\equiv&-b_{\pm\pm}\dl_\pm C^\pm
-\frac{1}{2}\dl_\pm b_{\pm\pm}C^\pm ~, \nn \\
T^{gh(3/2)}_{\pm\pm}&\equiv&\frac{3}{4}\be_{\pm\pm}\dl_\pm\w_\mp
+\frac{1}{4}\dl_\pm\be_{\pm\pm}\w_\mp ~, \nn \\
J^X_{\pm\pm}&\equiv&\CJ^X_\pm=\psi_\pm\dl_\pm X ~, \nn \\
J^L_{\pm\pm}&\equiv&\pm\frac{\k}{2}\Bigl[\h_\pm(\f'\pm\frac{2}{\k}\p_\f)
-4\h_\pm'-2\m e^{-\frac{\f}{2}}\h_\mp\Bigr] \nn \\
&=& \frac{\k}{2}(-\h_\pm\dl_\pm\f+2\dl_\pm\h_\pm) ~, \nn \\
J^{gh}_{\pm\pm}&\equiv&\mp i\Bigl(\frac{3}{4}\be_{\pm\pm}\dl_\pm C^\pm
+\frac{1}{2}\dl_\pm\be_{\pm\pm}C^\pm-4ib_{\pm\pm}\w_\mp\Bigr) ~,
\label{stsc}
\end{eqnarray}
where $\displaystyle{\p_\f\equiv\frac{\k}{2}\dot\f}$ is the canonical
momentum conjugate to $\f$. These results can be obtained directly from
(\ref{brsttr}) by using the equations of motion.

We now turn to the BRST charge $\tilde Q$ in the superconformal gauge
and examine the nilpotency. It is given by (\ref{QB}) with the
contribution of the BF fields included. Since $\CP^A=\be_z=0$ in this
gauge, we obtain
\begin{eqnarray}
\tilde Q&=&\int d\s \Bigl[C^+(T^X_{++}+T^L_{++}+\frac{1}{2}T^{gh(2)}_{++}
+T^{gh(3/2)}_{++})+C^-(T^X_{--}+T^L_{--}+\frac{1}{2}T^{gh(2)}_{--}
+T^{gh(3/2)}_{--}) \nn \\
&& -i\w_-(J_{++}^X+J_{++}^L)
+i\w_+(J_{--}^X+J_{--}^L)+2ib_{++}\w_-^2+2ib_{--}\w_+^2\Bigr] ~. \label{scQB}
\end{eqnarray}
As was mentioned in Section 3, the operator products in the rhs of
(\ref{scQB}) is defined by the ordering prescription. For the string
variables and the ghosts, it coincides with the free field normal
ordering. We do not know, however, what operator ordering should be
chosen for the BF fields. Since they are not free fields, the ordering
prescription introduced in Section 3 would not be completely legitimate.
But we shall continue to use the free field ordering prescription in
this section.

To satisfy the nilpotency of $\tilde Q$ it is necessary for the total
stress tensor and the total supercurrents defined by
\begin{eqnarray}
T^{\rm tot}_{\pm\pm}&=&T^X_{\pm\pm}+T^L_{\pm\pm}+T^{gh(2)}_{\pm\pm}
+T^{gh(3/2)}_{\pm\pm} ~, \nn \\
J^{\rm tot}_{\pm\pm}&=&J^X_{\pm\pm}+J^L_{\pm\pm}+J^{gh}_{\pm\pm} \label{Ttot}
\end{eqnarray}
to satisfy super-Virasoro algebra with total central charge being
canceled. Except for the BF sector, (\ref{stsc}) satisfy the
super-Virasoro algebra with central charges $3D/2$, $-26$, $11$ for
the string, the reparametrization ghost and the bosonic superghost
sectors, respectively. On the other hand the $T^L_{\pm\pm}$ and
$J^L_{\pm\pm}$ given in (\ref{stsc}) do not form super-Virasoro
commutation relations for nonvanishing cosmological term. To recover
the super-Virasoro algebra, let us modify these operators by
\begin{eqnarray}
T^L_{\pm\pm}&=&\frac{\k}{2}\Bigl[\frac{1}{4}(\f'\pm\frac{2}{\k}\p_\f)^2
-(\f'\pm\frac{2}{\k}\p_\f)'\pm\frac{i}{2}\h_\pm\h_\pm'+\m^2V_\al^2
\mp i\al\m V_\al\h_\mp\h_\pm\Bigr] ~, \nn \\
J^L_{\pm\pm}&=&\pm\frac{\k}{2}\Bigl[\h_\pm(\f'\pm\frac{2}{\k}\p_\f)
-4\h_\pm'-2\m V_\al\h_\mp\Bigr] ~, \label{mscsst}
\end{eqnarray}
where $\al$ is a free parameter and $V_\al$ is defined by
\begin{equation}
V_\al\equiv :\exp(-\al \f):~. \label{Vbe}
\end{equation}
For $\al=1/2$, (\ref{mscsst}) reduces to the classical expression
given in (\ref{stsc}). The operator products in the rhs of (\ref{mscsst})
are defined by the free field ordering prescription except for the
$V_\al^2$ term, while the operator $V_\al^2$ is assumed to be the
square of the normal ordered operators (\ref{Vbe}), hence it is not
well-defined. But as far as we know, it is not possible to recover
the super-Virasoro algebra by the free field ordering prescription
for nonvanishing cosmological term. Besides this shortcoming of
our argument, (\ref{mscsst}) satisfy the super-Virasoro algebra
with central charge $3/2+24\p\k$ for
\begin{equation}
\al\g-\frac{\al^2}{8\p}=\frac{1}{2}~. \label{beeq}
\end{equation}
This is just an extension of the canonical analysis of ref.
\cite{ct} to
superstring. The deviation of $\al$ from the classical value can
be interpreted
as gravitational self-dressing. The condition (\ref{beeq}) corresponds
to the requirement that the vertex operator (\ref{Vbe}) must be a
conformal field with conformal weight $(\frac{1}{2},\frac{1}{2})$
as discussed in ref. \cite{dk}.

These arguments show that the BRST charge (\ref{scQB}) satisfies the
nilpotency for $\k$ and $\al$ satisfying (\ref{mcoeff}) and
(\ref{beeq}), i.e.,
\begin{equation}
\al=\sqrt{\frac{\p}{2}}(\sqrt{9-D}\pm\sqrt{1-D}) ~. \label{beroot}
\end{equation}

\section{Supersymmetric light-cone gauge-fixing}
\setcounter{equation}{0}

In this section we will investigate the supersymmetric extension
of the light-cone gauge \cite{gx,pz,aaz}.
It is defined by the following set of
gauge conditions
\begin{equation}
e_+{}^+=e_-{}^-=1~, \qquad
e_-{}^+=0~, \qquad
\chi_{--}=\chi_{-+}=0 ~, \label{slcgc}
\end{equation}
on the zweibeins and gravitinos. These fix the gauges of
reparametrizations, local Lorentz and local supersymmetry.
To specify the gauge for the super-Weyl symmetry, we impose additional
gauge conditions on the super-Liouville fields as
\begin{equation}
\f=\h=F_L=0~.  \label{slcgcfslm}
\end{equation}
In this gauge the only independent components of the gravity multiplet
are $e_+{}^-\equiv -g_{++}$, $\chi_{++}$, $\chi_{+-}$ and $F_G$, which
behave as the analog of the super-Liouville fields.

The gauge conditions (\ref{slcgc}) and (\ref{slcgcfslm}) can be
implemented by choosing
\bn
&& \chi^{+}_\vp=N^{+}-1 ~,\quad \chi^{-}_\vp=(N^{-}+1)e^{\x}-2~,\quad
\chi^{\ve}=\ve ~, \quad \chi^{\x}=\f ~, \quad \chi^f=f_L ~,\nn \\
&& \chi^{+}_\CJ=-iM^{+} ~, \quad
\chi^{-}_\CJ=iM^{-}-\frac{1}{4}(N^-+1)\Lam_- ~, \quad
\chi^{\pm}_{\Lam}=\h_{\pm} ~.
\label{slcgcd}
\en
It can be easily seen that these together with $N^\pm=\lam^\pm$ and $M^\pm=
\pm i\n_\mp$ given by (\ref{gcond}) indeed lead to (\ref{slcgc}) and
(\ref{slcgcfslm}.

The gauge conditions
(\ref{slcgcd}) do not satisfy the assumption given below (\ref{gcond})
since their BRST transforms contain the canonical momentum $\p_\th$ and
$\z_\pm$. Then
the master action (\ref{gfa}) formulated in the EPS leads to erroneous
results when (\ref{slcgcd}) are imposed. This is because (\ref{Liouv})
can not be identified with the covariant super-Liouville fields. It seems
rather difficult to find covariant super-Liouville multiplet without the
assumption on the gauge conditions. To avoid this difficulty, we apply
(\ref{slcgcd}) for (\ref{gfca}) instead of (\ref{gfa}) and use the
covariant BRST transformations (\ref{ncbrsttr}) rather than (\ref{brsttr}).
This should be compared with the superconformal gauge-fixing where the
gauge conditions (\ref{scgcd}) satisfy the assumption
given below (\ref{gcond}),\footnote{In the case of EPS the gauge condition
corresponding to $f_L=0$ is inherently absent.} and we can work with
(\ref{gfa}) from the beginning as well.

Substituting (\ref{slcgcd}) into (\ref{SLetc}), we obtain the effective
action (\ref{gfca}) in the light-cone gauge.\footnote{In this section
we consider the case $\m^2=0$ for simplicity.} As in the superconformal
gauge-fixing, $C_L$, $C_W$, $\h_{\scriptscriptstyle W}$ and $F_W$ in the
ghost sector become nonpropagating, and can be eliminated via equations
of motion. Taking variations of $S_{\rm eff}$ with respect to $N^\pm
=\lam^\pm$ and $M^\pm=\pm i\n_\m$, we obtain the multipliers $B_\pm$
and $A^\pm$ as
\begin{eqnarray}
&& B_+=\vp^X_++\vp^g_++\vp^{gh}_+ ~, \nn \\
&& g_{11}B_-=\vp^X_-+\vp^g_-+\vp^{gh}_-
+\frac{i}{4}(\CJ^X_-+\CJ^g_-+\CJ^{gh}_-)\Lam_-~, \nn \\
&& A^\pm=\pm i(\CJ^X_\mp+\CJ^g_\mp+\CJ^{gh}_\mp) ~, \label{lcauxf}
\end{eqnarray}
where $\vp^{X,g,gh}_\pm$ and $\CJ^{X,g,gh}_\pm$ are the super-Virasoro
constraints given by
\begin{equation}
\vp^{X,g,gh}_\pm\equiv-\frac{\de S_{X,g,{\rm gh}}}{\de \lam^\pm} ~,
\qquad
\CJ^{X,g,gh}_\pm\equiv\mp i\frac{\de S_{X,g,{\rm gh}}}{\de \n_\mp} ~.
\label{lcsvcons}
\end{equation}
We can relate the BRST transformations of the anti-ghosts with the
super-Virasoro operators via (\ref{lcauxf}).

In the light-cone gauge, the ghost action takes the following form
\bn
S_{\rm gh}&=&\int d^2 \s\biggl[
-{\overline{\cal C}}_+\dl_-C^+-{\overline{\cal C}}_-
(\dl_+C^++\dl_-C^--g_{11}\dl_-C^+-2i\w_-\Lam_+) \nn \\
&&+{\overline{\g}}_+\dl_-\w_-+{\overline{\g}}_-
\biggl(\dl_-\w_++\frac{1}{2}\w_+\dl_-\x+\frac{1}{4}\dl_-C^+\Lam_-
-\frac{1}{2\sqrt{g_{11}}}f_G\w_-\biggr)\biggr]
\label{slcgfa}
\en
The presence of the auxiliary field $f_G$ turns out to be essential
in order for the BRST transformations to be consistent with
the equations of motion. Eliminating $f_G$ via the equation of motion
$\k\sqrt{g_{11}} f_G=\overline\g_-\w_-$, we see that there appears
a quartic term in the bosonic ghosts and anti-ghosts. The
$S_{\rm gh}$ also contains interaction terms between the supergravity sector
and the ghost sector. Remarkably enough, all the ghost variables and
$\Lam_+$ can be made free fields by the following field redefinitions
\bn
c^+ &\equiv& C^+ \nn \\
c_+ &\equiv& C^-+\frac{x^-}{2}\dl_+C^+-ix^-\w_-\Lam_++\frac{i}{k}
(x^-)^2{\overline{\cal C}}_-(\w_-)^2 \nn\\
\g_- &\equiv& \w_- \nn\\
\g_+ &\equiv& \sqrt{g_{11}}\w_+-\frac{x^-}{4k}\frac{{\overline\g}_-}
{\sqrt{g_{11}}}(\w_-)^2 \label{newghost}\\
b &\equiv& {\overline{\cal C}}_- \nn\\
b_{++} &\equiv& {\overline{\cal C}}_+-g_{11}{\overline{\cal C}}_-
+\frac{1}{4}{\overline\g}_-\Lam_-+\frac{x^-}{2}\dl_+{\overline{\cal
C}}_- \nn \\
\be_+ &\equiv& \frac{{\overline\g}_-}{\sqrt{g_{11}}} \nn\\
\be_{++} &\equiv& {\overline\g}_+-ix^-{\overline{\cal C}}_-\Lam_+
+\frac{x^-}{4k}\biggl(\frac{{\overline\g}_-}
{\sqrt{g_{11}}}\biggr)^2\w_- ~, \nn \\
\chi_+ &\equiv& \Lam_+-\frac{2}{k}x^-{\overline{\cal C}}_-\w_- ~. \nn
\en
In terms of these variables the gauge-fixed action then takes the form
\begin{eqnarray}
S_{\rm eff}&=&\int d^2x \Bigl[~\frac{1}{2}\dl_-X(\dl_+X+g_{++}\dl_-X)
+\frac{i}{2}\psi_+\dl_-\psi_+ \nn \\
&&+\frac{i}{2}\psi_-(\dl_+\psi_-+g_{++}\dl_-\psi_-)
-2i\chi_{++}\psi_-\dl_-X\Bigr] \nn \\
&& +\frac{\k}{2}\int d^2x
\Bigl[~\frac{1}{2g_{11}}\{(\dl_-g_{11})^2-2\dl_-g_{11}
(\ln g_{11})'+4(\ln g_{11})''\} \nn \\
&&+\frac{8i}{g_{11}}\chi_{++}(\dl_-\chi_{++}
-\frac{2}{g_{11}}\chi_{++}')+\frac{i}{2}\chi_+\dl_-\chi_+\Bigr] \nn \\
&& +\int d^2x(-b_{++}\dl_-c^+-b\dl_-c_++\be_{++}\dl_-\g_-
+\be_{+}\dl_-\g_+)~, \label{lcgfseff}
\end{eqnarray}
where we have used the relations $g_{11}=1+g_{++}$ and $\sqrt{g_{11}}
\Lam_-=4\chi_{++}$, and $\psi_+/\sqrt{g_{11}}$, the lower component of
fermionic string coordinates $\psi$ in our representation, has been newly
denoted by $\psi_+$. The $S_{\rm eff}$ is the supersymmetric extension
of 2D gravity action discussed in \cite{ftimk}. As in the bosonic string
case, it leads to the free ghost equations
\bn
&& \dl_-c^+=\dl_-c_-=\dl_-\g_-=\dl_-\g_+=0 ~, \nn\\
&& \dl_-b_{++}=\dl_-b=\dl_-\be_{++}=\dl_-\be_+=0 ~,
\label{freeghost}
\end{eqnarray}
and the canonical supercommutation relations among ghost variables as
\begin{eqnarray}
&& [c^+(\s),b_{++}(\s')]=[c_+(\s),b(\s')]=-i\de(\s-\s') ~, \nn \\
&& [\g_-(\s),\be_{++}(\s')]=[\g_{+}(\s),\be_+(\s')]=i\de(\s-\s') ~,
\label{ghostCCR} \\
&& {\rm all~other~supercommutators~vanish}~. \nn
\end{eqnarray}
By taking variations of (\ref{lcgfseff}) with respect to $g_{++}$ and
$\chi_{++}$, we obtain the equations of motion for graviton and gravitino
as
\begin{eqnarray}
&& \vp_-^X+\vp_-^g=\frac{\k}{4}g_{11}\dl_-^2g_{11}
-4i\k\chi_{++}\dl_-\chi_{++} ~, \nn \\
&& \CJ_-^X+\CJ_-^g=4\k\sqrt{g_{11}}\dl_-\chi_{++} ~,\label{WTid}
\end{eqnarray}
where we have used (\ref{lcsvcons}).

The BRST transformations of the variables appearing in (\ref{lcgfseff})
can be found from (\ref{newghost}) and (\ref{ncbrsttr}), by using
the gauge conditions (\ref{slcgc}) and the equations of motion for ghosts
(\ref{freeghost}) as follows
\bn
\de g_{++}&=&\frac{1}{2}c^+\dl_+ g_{++}+g_{++}\dl_+ c^+
+\frac{1}{2}\Bigl(c_+-\frac{x^-}{2}\dl_+ c^+\Bigr)\dl_- g_{++}
-\frac{1}{2}\dl_+\Bigl(c_+-\frac{x^-}{2}\dl_+ c^+\Bigr) \nn \\
&& +4i\g_+\chi_{++}-i\g_-\chi_+\Bigl(1-\frac{x^-}{2}\dl_-\Bigr)g_{++}
-\frac{i}{2}x^-\dl_+(\g_-\chi_+) \nn \\
&& +\frac{i}{\k}x^-\be_-\g_-^2\chi_{++}
-\frac{2i}{\k}b\g_-^2\Bigl(x^--\frac{(x^-)^2}{4}\Bigr)g_{++}
-\frac{i}{2\k}(x^-)^2\dl_+(b\g_-^2) ~, \nn \\
\de\chi_{++}&=&\frac{1}{2}c^+\dl_+\chi_{++}+\frac{3}{4}\dl_+ c^+\chi_{++}
+\frac{1}{2}\Bigl(c_+-\frac{x^-}{2}\dl_+\Bigr)\dl_-\chi_{++} \nn \\
&& -\frac{i}{2}\g_-\chi_+(1-x^-\dl_-)\chi_{++}
-\frac{i}{\k}b\g_-^2\Bigl\{x^--\frac{(x^-)^2}{2}\dl_-\Bigr\}\chi_{++} \nn \\
&& -\frac{1}{4}\g_+\dl_- g_{++}+\frac{1}{4\k}\be_+\g_-^2
\Bigl(1-\frac{x^-}{4}\dl_-\Bigr)g_{++}
+\frac{x^-}{8\k}\dl_+(\be_+\g_-^2) ~, \nn \\
\de \chi_+&=&\frac{1}{2}c^+\dl_+\chi_+-\frac{1}{4}\dl_+ c^+\chi_+
+2\dl_+\g_- \nn \\
&& +\g_-\Bigl(1-\frac{x^-}{2}\dl_-\Bigr)\dl_- g_{++}
-\frac{2}{\k}\g_-\Bigl(bc_+-\frac{1}{2}\be_+\g_+\Bigr) ~, \nn \\
\de c^+&=&\frac{1}{2}c^+\dl_+ c^++2i\g_-^2 ~, \nn \\
\de c_+&=&\frac{1}{2}c^+\dl_+ c_++\frac{1}{2}\dl_+ c^+c_++2i\g_+^2 ~, \nn \\
\de\g_-&=&\frac{1}{2}c^+\dl_+\g_--\frac{1}{4}\dl_+ c^+\g_- ~, \nn \\
\de\g_+&=&\frac{1}{2}c^+\dl_+\g_++\frac{1}{4}\dl_+ c^+\g_+
-\frac{i}{2}\g_+\g_-\chi_+ ~, \nn \\
&&-2i\g_-^2\Bigl(1-\frac{x^-}{2}\dl_-\Bigr)\chi_{++}
+\frac{1}{4\k}c_+\be_+\g^2_- \nn \\
\de b_{++}&=&
T_{++}^{X}+T_{++}^{g}+\frac{1}{2}T_{++}^{gh(2)}+T_{++}^{gh(0)}
+T_{++}^{gh(3/2)}
+T_{++}^{gh(1/2)} ~,
\label{slcbrst}\\
\de b&=&\frac{\k}{4}\dl_-^2g_{++}-\frac{1}{2}\dl_+ bc^+
-ib\g_-\chi_++\frac{1}{8\k}(\be_+\g_-)^2 ~, \nn \\
\de \be_{++}&=&iJ^X_{++}+\frac{i\k}{2}\chi_+
\Bigl(1-\frac{x^-}{2}\dl_-\Bigr)\dl_- g_{++}-i\k\dl_+\chi_+ \nn \\
&& +\frac{1}{2}\dl_+\be_{++}c^++\frac{3}{4}\be_{++}\dl_+ c^+
-4ib_{++}\g_- \nn \\
&& -4ib\g_-\{1-\frac{x^-}{2}\dl_-+\frac{(x^-)^2}{8}\dl_-^2\}g_{++}
+4i\be_+\g_-\Bigl(1-\frac{x^-}{2}\dl_-\Bigr)\chi_{++} \nn \\
&& -i\bigl(bc_+-\frac{1}{2}\be_+\g_+\bigr)\chi_+
-\frac{1}{4\k}\be_+^2\g_-c_+ ~, \nn \\
\de\be_+&=&-4i\k\dl_-\chi_{++}+\frac{1}{2}\dl_+\be c^+
+\frac{1}{4}\be_+\dl_+ c^+-4ib\g_++\frac{i}{2}\be_+\g_-\chi_+ ~, \nn
\en
where we have omitted the transformations of string variables. In deriving
the BRST transformations of anti-ghosts, we have used (\ref{lcauxf}) and
(\ref{WTid}). The stress tensors and the supercurrents are defined by
\bn
T_{++}^X&\equiv& \vp^X_+
=\frac{1}{4}[(\dl_+X+g_{++}\dl_-X)^2+i\psi_+\dl_+\psi_+] \nn\\
J_{++}^X&\equiv & \CJ^X_+=\psi_+(\dl_+X+g_{++}\dl_-X) ~, \nn \\
T_{++}^{g}&\equiv& \frac{\k}{2}\biggl[~\frac{1}{4}(\dl_-g_{++})^2
-\frac{1}{2}g_{++}\dl_-^2g_{++}
-\frac{1}{2}(\dl_--\frac{x^-}{2}\dl_-\dl_+)\dl_-g_{++}\biggr]  \nn\\
& & +4i\k\chi_{++}\dl_-\chi_{++}+\frac{i\k}{8}\chi_+\dl_+\chi_+ ~, \nn \\
T_{++}^{gh(2)}&\equiv & -\frac{1}{2}\dl_+b_{++}-b_{++}\dl_+c^+ ~,
\label{stresstensor}\\
T_{++}^{gh(0)}&\equiv& \frac{1}{2}\dl_+bc^+ ~, \nn \\
T_{++}^{gh(3/2)}&\equiv &\frac{3}{4}\be_{++}\dl_+\g_-
+\frac{1}{4}\dl_+\be_{++}\g_- ~, \nn \\
T_{++}^{gh(1/2)}&\equiv&\frac{1}{4}\be_+\dl_+\g_+
-\frac{1}{4}\dl_+\be_+\g_+ ~. \nn
\en
Except for the gravitational stress tensor $T^g_{++}$, these satisfy
conservation laws. The index $j=2,0,3/2,1/2$ of the stress tensors for the
ghost sector labels the canonical pairs of ghost and anti-ghost with
conformal weights $1-j$ and $j$  \cite{fms}. The $T^{gh(j)}_{++}$ defined by
the usual normal ordering satisfies the Virasoro algebra with central charge
\begin{equation}
2\e(6j^2-6j+1) ~, \label{cghj}
\end{equation}
where $\e$ stands for the grassmannian parity of the ghost pair. The total
central charge of the ghost sectors is then given by
\begin{equation}
c_{gh}=-18 ~. \label{cghtot}
\end{equation}

Using (\ref{ghostCCR}) and the BRST transformations for the ghosts, we can
construct the BRST charge generating (\ref{slcbrst}) as
\bn
\tilde Q&=&\int
d\s\Biggl[~c^+\biggl(T_{++}^X+T_{++}^{g}+\frac{1}{2}T_{++}^{gh(2)}
+T_{++}^{gh(0)}+T_{++}^{gh(3/2)}+T_{++}^{gh(1/2)}\biggr) \nn\\
&&+c_+\biggl(\frac{\k}{4}\dl_-^2g_{++}-ib\g_-\chi_+
+\frac{1}{8\k }(\be_+\g_-)^2\biggr) \nn\\
&&+\g_-\biggl(-iJ^X_{++}
-\frac{i\k}{2}\chi_+\biggl(1-\frac{x^-}{2}\dl_-\biggr)\dl_-g_{++}
+i\k\partial_+\chi_++2ib_{++}\g_- \nn \\
&& +2ib\g_-\biggl(1-\frac{x^-}{2}\dl_-+\frac{(x^-)^2}{8}\dl^2_-\biggr)g_{++}
-2i\be_+\g_-\biggl(1-\frac{x^-}{2}\dl_-\biggr)\chi_{++}\biggr) \nn \\
&&+\g_+\biggl(4i\k\dl_-\chi_{++}+2ib\g_+
-\frac{i}{2}\be_+\g_-\chi_+\biggr)\Biggr] ~,
\label{slcbrstcharg}
\en
where normal products among the ghost variables are implicitly assumed.

The BRST invariance of (\ref{lcgfseff}) implies that (\ref{slcbrst})
must be consistent with the equations of motion (\ref{freeghost}). We thus
obtain the supercurvature equations
\begin{equation}
\dl_-^3g_{++}=0~, \qquad \dl_-^2\chi_{++}=0 ~, \qquad
\dl_-\chi_+=0 ~. \label{sceq}
\end{equation}
Then the gravitational stress tensor $T_{++}^g$ in (\ref{stresstensor})
also turns out to be conserved, i.e.,
\begin{equation}
\dl_-T^g_{++}=0 ~. \label{Tgcons}
\end{equation}
\setcounter{footnote}{0}
By virtue of (\ref{freeghost}), (\ref{sceq}) and the conservation
of $T^X_{++}$ and $J^X_{++}$,\footnote{The possible anomalies in these
currents due to the super-Virasoro anomaly indeed vanish in the light-cone
gauge where $\lam^+=1$ and $\nu_-=0$ \cite{fs}.} we can show that the BRST
charge (\ref{slcbrstcharg}) is a constant of motion.

It remains to show the nilpotency of $\tilde Q$.
Before turning to this issue, we must fix the commutation relations for
the 2D supergravity sector. This can be done by comparing the BRST
transformations (\ref{slcbrst}) and (\ref{tilBRST}) for $g_{++}$ and
$\chi_{++}$. We first expand these operators in terms of conserved
currents by noting (\ref{sceq}) as
\bn
g_{++} &=& -\frac{1}{2\kappa}[J^{+}(x^{+})-2x^{-}
J^{0}(x^{+})+(x^{-})^2 J^{-}(x^{+})] ~,\nn\\
\chi_{++}&=& -\frac{1}{2\kappa}[\Psi^{-1/2}
(x^+)+x^-\Psi^{1/2}(x^+)] ~.\label{g++}
\en
The BRST transformations of $g_{++}$ and $\chi_{++}$ given in
(\ref{slcbrst}) can be transcribed into the transformations of
these currents as
\bn
\de J^+ &=& \frac{1}{2}c^+ \partial_+ J^++\partial_+ c^+
 J^+-2c_+ J^{0}+\k\partial_+ c_++4i\g_+\Psi^{-1/2}
-i\g_-\chi_+J^+ ~, \nn \\
\de J^{0} &=& \frac{1}{2}c^+\partial_+J^{0}
+\frac{1}{2}\partial_+c^+J^0-c_+ J^-+\frac{\k }{4}\partial^2_+ c^+
-2i\g_+\Psi^{1/2} \nn\\
&& -\frac{i\k}{2}\dl_+(\g_-\chi_+)+\frac{i}{\k}(\g_-)^2\Bigl(bJ^+
-\frac{1}{2}\be_+\Psi^{-1/2}\Bigr) ~, \nn\\
\de J^{-} &=& \frac{1}{2}c^+\partial_+J^{-}+i\g_-\chi_+J^-
+\frac{2i}{\k}(\g_-)^2\Bigl(bJ^0+\frac{1}{2}\be_+\Psi^{1/2}\Bigr)
+i\dl_+(b(\g_-)^2) ~,  \nn \\
\de \Psi^{-1/2}&=&
\frac{1}{2}c^+\dl_+\Psi^{-1/2}+\frac{3}{4}\dl_+c^+\Psi^{-1/2}
+c_+\Psi^{1/2} \label{brstJ} \\
&& -\k\dl_+\g_++\g_+J^0-\frac{i}{2}\g_-\chi_+\Psi^{-1/2}
+\frac{1}{4\k }\be_+(\g_-)^2J^+ ~,\nn\\
\de\Psi^{1/2}&=&\frac{1}{2}c^+\dl_+\Psi^{1/2}
+\frac{1}{4}\dl_+c^+\Psi^{1/2}-\g_+J^-
+\frac{i}{2}\g_-\chi_+\Psi^{1/2} \nn \\
&& -\frac{1}{4\k}(\g_-)^2\Bigl(\be_+J^0+4ib\Psi^{-1/2}\Bigr)
-\frac{1}{4}\dl_+(\be_+(\g_-)^2) ~, \nn \\
\de\chi_+&=&\frac{1}{2}c^+\dl_+\chi_++\frac{1}{4}\dl_+c^+\chi_+
+2\dl_+\g_-+\frac{2}{\k}\g_-\Bigl(J^0-bc_++\frac{1}{2}\be_+\g_+\Bigr) ~. \nn
\en
Then $J^a$ and $\Psi^r$ can be shown to satisfy OSp(1,2) Kac-Moody
current algebra
\bn
[~J^{a}(\s)~,~J^{b}(\s')~] &=& if^{ab}{}_{c}~J^{c}(\s)~
\delta (\s - \s') - i\k \eta^{ab}
\dl_\s\de(\s - \s') ~, \nn \\
{}[~J^{a}(\s)~,~\Psi^{r}(\s')~] &=& if^{ar}{}_{s}~\Psi^{s}(\s)~
\delta (\s - \s') ~, \label{sl2rqu}\\
{}[~\Psi^{r}(\s)~,~\Psi^{s}(\s')~] &=& f^{rs}{}_{a}~J^{a}(\s)~
\delta (\s - \s')-\k\eta^{rs}
\dl_\s\de(\s - \s')~, \nn
\en
where $f$'s and $\h$'s are, respectively, the structure constants and the
Killing metric for OSp(1,2) algebra. They satisfy $f^{ab}{}_c
=-f^{ba}{}_c$, $f^{rs}{}_a=f^{sr}{}_a$, $\h^{ab}=\h^{ba}$ and
$\h^{rs}=-\h^{sr}$ with nonvanishing components $f^{\pm0}{}_\pm=\pm1$,
$f^{+-}{}_0=2$, $f^{+~1/2}{}_{-1/2}=-f^{-~-1/2}{}_{1/2}=-1$,
$f^{0~1/2}{}_{1/2}=-f^{0~-1/2}{}_{-1/2}=1/2$, $f^{-1/2~-1/2}{}_+=
-f^{-1/2~1/2}{}_0=f^{1/2~1/2}{}_-=-1/4$, $\h^{+-}=-2 \h^{00}=2$ and
$\h^{-1/2~1/2}=1/2$.

It is natural to redefine operator ordering for the gravitational sector
to ensure the symmetry associated with the OSp(1,2) current algebra.
To this end we decompose $J^a(x^+)$ into positive and negative frequency
parts by
\begin{eqnarray}
J^{a(\pm)}(x^+)=\int dy^+\de^{(\mp)}(x^+-y^+)J^a(y^+) ~, \label{decompJ}
\end{eqnarray}
and similarly for $\Psi^r(x^+)$. We then define operator ordering with
respect to this decomposition. The gravitational stress
tensor $T^g_{++}$ given in (\ref{stresstensor}) must be defined by the
Sugawara form
\begin{equation}
T^g_{++}=-\frac{1}{2\k'}:(\h_{ab}J^aJ^b+i\h_{rs}\Psi^r\Psi^s):
-\frac{1}{2}\dl_+J^0+\frac{i\k'}{8}\chi_+\dl_+\chi_+ ~, \label{qTg}
\end{equation}
where $\h_{ab}$ and $\h_{rs}$ are the inverses of $\h^{ab}$ and $\h^{rs}$.
The parameter $\k'$ is modified from its classical value $\k$ by
\begin{equation}
\k'=\k-\frac{3}{8\p} ~. \label{kprime}
\end{equation}
The stress tensor thus defined not only ensures the BRST transformations
(\ref{brstJ}) but also satisfies the Virasoro algebra with the central
charge given by
\begin{equation}
c_g\equiv\frac{2k}{2k-3}+6k+\frac{1}{2} ~, \label{cg}
\end{equation}
where $k\equiv4\p\k$ is the central charge of the current algebra
(\ref{sl2rqu}). The last term in the rhs of (\ref{cg}) is the
contributions due to $\chi_+$.

In (\ref{qTg}) we have used the rescaled $\chi_+$ by
$\displaystyle{\sqrt{\frac{\k}{\k'}}\chi_+\rightarrow\chi_+}$ satisfying
\begin{equation}
[\chi_+(x^+),\chi_+(y^+)]=\frac{2}{\k'}\de(x^+-y^+) ~. \label{chichi}
\end{equation}
The stress tensor (\ref{qTg}) can be regarded as that given in
(\ref{stresstensor}) with all the parameter $\k$ replaced by $\k'$.
This enables one to interpret that the quantum modifications appears
not in the stress tensor but in the current algebra (\ref{sl2rqu}), hence
in (\ref{brstJ}). Since $\k$ is a free parameter both interpretation can
be legitimated.

We are now in a position to investigate the nilpotency of the BRST charge.
The $\tilde Q$ given by (\ref{slcbrstcharg}) does not satisfy the
nilpotency even after the substitution of (\ref{qTg}) for $T^g_{++}$.
We must replace $\k$ appearing in (\ref{slcbrstcharg}) by $\k'$
corresponding to the change (\ref{kprime}). After a rather lengthy
computations, it can be shown that the BRST charge thus defined
yet contains a BRST anomaly of trivial type as well as the nontrivial
one \cite{fikkt}. The former can be removed by shifting the BRST charge by
\begin{equation}
\frac{i}{8\p}\int d\s\g_-\dl_+\chi_+ ~. \label{cobQ}
\end{equation}
The correct quantum mechanical BRST charge is finally given by
\bn
\tilde Q&=&\int
d\s:\Biggl[~c^+\biggl(T_{++}^X+T_{++}^{g}+\frac{1}{2}T_{++}^{gh(2)}
+T_{++}^{gh(0)}+T_{++}^{gh(3/2)}+T_{++}^{gh(1/2)}\biggr) \nn\\
&&+c_+\biggl(-J^--ib\g_-\chi_++\frac{1}{8\k'}(\be_+\g_-)^2\biggr) \nn\\
&&+\g_-\biggl(-iJ^X_{++}-i\chi_+J^0+i\biggl(\k'+\frac{1}{8\p}\biggr)
\partial_+\chi_++2ib_{++}\g_-
-\frac{i}{\k'}\g_-(bJ^+-\be_+\Psi^{-1/2})\biggr) \nn \\
&&+\g_+\biggl(-4i\Psi^{1/2}+2ib\g_+
-\frac{i}{2}\be_+\g_-\chi_+\biggr)\Biggr]: ~.
\label{qslcbrstcharg}
\en
The $\tilde Q^2$ contains only the cohomologically nontrivial anomaly
of the type (\ref{solution}) given by
\begin{equation}
\tilde Q^2=-\frac{ic_{\rm tot}}{48\p}\int d\s
(c^+c^+{}'''+8i\g_-\g_-'') ~, \label{qQ2}
\end{equation}
where $c_{\rm tot}=c_X+c_g+c_{gh}$ is the total Virasoro central charge.
We thus arrive at the KPZ condition for the nilpotent BRST charge in the
$N=1$ NSR superstring as
\begin{equation}
c_{\rm tot}=\frac{3}{2}D+\frac{2k}{2k-3}+6k+\frac{1}{2}-18=0 ~,
\label{totalcc}
\end{equation}
where use has been made of the results (\ref{cghtot}) and (\ref{cg}) as
well as $c_X=\frac{3}{2}D$ for the string sector.

\section{Summary and discussion}

We have investigated BRST quantization of the NSR superstring at
noncritical dimensions as 2D SUGRA coupled with the string
variables. It is done with special emphasis on the point that the
super-Liouville mode which is decoupled from the theory at the
classical level becomes dynamically active through the
superconformal anomaly. At noncritical dimensions the
super-Virasoro anomaly destroying the reparametrization
invariance and the local supersymmetry can be canceled by
introducing the BF fields. This naturally leads to a
gauge symmetric extension of the original system which
suffers from the super-Virasoro anomaly.

The gauge-fixed effective action thus obtained turns out to contain two
actions of super-Liouville type. The one written only in terms of
2D supergravity fields can be regarded as the counterterm
removing the super-Virasoro anomaly. This action turns out
to reproduce correct super-Weyl anomaly as was argued in
\cite{fikkt}. The BF fields constitute the other one, which
cancels the super-Weyl anomaly. By introducing the BF fields
we have been able not only to construct an effective action
possessing all the classical local symmetries but also show
within canonical formalism how the super-Liouville mode acquires
dynamical behavior through the superconformal anomaly without
invoking particular gauge conditions or weak field approximations.

As we have mentioned in Section 4, it is possible to gauge-fix
the 2D supergravity fields to flat ones as in refs. \cite{imnu}.
These authors ignored the superconformal anomaly, and necessarily
found that the theory is only consistent at the critical dimensions
\cite{ko,imnu}. In the present case, the BRST invariance does not
leads to any inconsistency even at noncritical dimensions but
yields the vanishing condition of the total central charge and
the gravitational dressing effect \cite{dk} in the superconformal
gauge. The well-known barrier at $D=1$ also arises in our approach
as can be seen from (\ref{beroot}), indicating the breakdown of
validity of the continuum Liouville approach \cite{gm}. The
superconformal mode does
not decouple from the theory and is described by the supersymmetric
extension of the Liouville action given by DDK \cite{d,dk}. By simply
transcribing our canonical argument into path integral one, the
functional measure for the super-Liouville mode turns out to be
translational invariant.
This provides a canonical verification of
the functional measure ansatz of DDK. This can be understood as
follows. The essential point that
leads the authors of \cite{dk} to their super-Liouville action is the
fake super-Weyl invariance arising in the decompositions of
2D supergravity fields into super-Liouville mode and fiducial
background fields. Requiring the symmetry not to be broken by
superconformal anomaly necessarily results in the anomaly-canceling
super-Liouville action given in \ref{gfascg} up to trivial rescaling
of the fields. Since canceling the
BRST anomaly in our BFV-BRST approach is equivalent to eliminating
superconformal anomaly, we arrive at the effective action of ref.
\cite{dk}.

One of the advantageous point of our canonical approach is that
the effective action (\ref{gfca}) is a local functional without
referring to any particular gauge and arbitrary gauges can be
argued on an equal footing. In particular, we can explain the
manifestation of OSp(1,2) current algebra from the BRST invariance
in the light-cone gauge. This is contrasted with the approaches of
ref. \cite{gx,pz,aaz,kura}. These authors started with the anomalous
Ward-Takahashi identities corresponding to the supercurvature equations
(\ref{sceq}) and then extracted OSp(1,2) current algebra. In ref.
\cite{kura,ih}, BRST analyses were carried out for the system with the
stress tensor and the supercurrent obtained by applying Sugawara
construction. Although the KPZ condition (\ref{totalcc}) coincides
with the result of ref. \cite{pz,kura} for N=1 2D SUGRA,
there are crucial differences between ours and that in ref.\cite{pz,kura} in
the ghost content and, consequently,
the expression of the BRST charge (\ref{qslcbrstcharg}). In ref.
\cite{kura}, six ghost-antighost pairs were introduced corresponding
to the six generators of the residual transformations leaving the
light-cone gauge unchanged. In our case, there are four pairs and
the rest can be eliminated by the equations of motion as multiplier
fields. What yields these qualitative differences is the inclusion of
the auxiliary fields to ensure the off-shell nilpotency of the BRST
transformations (\ref{cbrsttr}) and their appearance in the effective
action (\ref{gfca}) through the super-Weyl anomaly. This leads to the
nontrivial redefinition of $\chi_{+-}$ as in (\ref{newghost}) and the
ghost higher order terms in (\ref{qslcbrstcharg}), to which very little
attention seems to have been payed so far.
In this paper the inclusion of the supersymmetric auxiliary fields
has been done only after passing to the configuration space.
Systematic methods for including such variables in EPS seems to be
lacking yet, and it is certainly worth exploring them for the BFV-BRST
formalism.

\vskip .5cm
\noindent{\bf Acknowledgments}

T. F. is supported in part by the Grant-in-Aid for Scientific
Research (No. 06740198) from the Ministry of Education, Science
and Culture. Y. I. would like to thank YITP for hospitality.

\appendix
\section{BRST transformations}
\setcounter{equation}{0}

In this appendix we summarize the BRST transformations in the
configuration space. The covariant ghosts are denote by
$C^\al$, $C_L$, $\w$, $C_W$ and $\h_{\scriptscriptstyle W}$ for
reparametrizations, local Lorentz, local supersymmetry,
Weyl rescaling and fermionic transformations, respectively.
In addition to these, we introduce the auxiliary fields
$F_X^\mu$, $F_G$, $F_L$ and $F_W$ for supermultiplets of
string variables, 2D supergravity, super-Liouville fields and
Weyl ghosts, respectively, to ensure the off-shell nilpotency
of the BRST transformations. The complete list of them is given by
\begin{eqnarray}
&& \de X=C^\al\partial_\al X+\overline\w\psi ~, \nn \\
&& \de \psi=-\frac{1}{4}C_W\psi+C^\al\partial_\al\psi+\frac{1}{2}
C_L\r_5\psi
-i\r^\al\w(\partial_\al X-\overline\chi_\al\psi)+\w F_X ~, \nn \\
&& \de F_X=-\frac{1}{2}C_WF_X+C^\al\partial_\al F_X
-i\overline\w\r^\al
[\nabla_\al\psi+i\r^\be(\partial_\be X-\overline\chi_\be\psi)\chi_\al
-\chi_\al F_X] ~,\nn \\
&& \de\f=C_W+C^\al\partial_\al\f+\overline\w\h ~, \nn \\
&& \de\h=-\frac{1}{4}C_W\h-\h_{\scriptscriptstyle
W}+C^\al\partial_\al\h+\frac{1}{2}
C_L\r_5\h-i\r^\al\w(\partial_\al\f-\overline\chi_\al\h)
+\w F_L ~, \nn \\
&& \de F_L=-\frac{1}{2}C_W F_L+F_W+C^\al\partial_\al F_L
-i\overline\w\r^\al[\nabla_\al\h+i\r^\be(\partial_\be\f
-\overline\chi_\be\h)\chi_\al-\chi_\al F_L] ~, \nn \\
&& \de e_\al{}^a=\frac{1}{2}C_We_\al{}^a
+C^\be\partial_\be e_\al{}^a+\partial_\al C^\be e_\be{}^a
+\e^a{}_bC_Le_\al{}^b-2i\overline\w\r^a\chi_\al~,
\nn \\
&& \de\chi_\al=\frac{1}{4}C_W\chi_\al+\frac{i}{4}\r_\al
\h_{\scriptscriptstyle W}+C^\be\partial_\be\chi_\al
+\partial_\al C^\be\chi_\be+\frac{1}{2}C_L\r_5\chi_\al+\nabla_\al\w
-\frac{i}{4}\r_\al\w F_G ~, \nn \\
&& \de F_G=-\frac{1}{2}C_WF_G+F_W+C^\al\partial_\al F_G \nn \\
&& ~~~~~~~~~~-4\e^{\al\be}\overline\w\r_5[\nabla_\al\chi_\be
+i(\overline\chi_\al\r_\be\chi^\g+\overline\chi_\be\r_\al\chi^\g
+\overline\chi_\al\r^\g\chi_\be)\chi_\g]
+i\overline\w\r^\al\chi_\al F_G ~, \nn \\
&& \de C^\al=C^\be\partial_\be C^\al+i\overline\w\r^\al\w ~, \nn \\
&& \de\w=\frac{1}{4}C_W\w+C^\al\partial_\al\w+\frac{1}{2}C_L\r_5\w
-i\overline\r^\al\w\chi_\al ~, \nn \\
&& \de C_L=C^\al\dl_\al C_L+\frac{1}{2}F_G\overline\w\r_5\w
-\frac{1}{2}\overline\w\r_5\h_{\scriptscriptstyle W}
-i\e_{ab}e^{\be a}\overline\w\r^\al\w(\dl_\al e_\be{}^b
-\dl_\be e_\al{}^b-i\overline\chi_\al\r^b\chi_\be) ~, \nn \\
&& \de C_W=C^\al\partial_\al C_W+\overline\w\h_{\scriptscriptstyle W} ~, \nn \\
&&\de\h_{\scriptscriptstyle W}=-\frac{1}{4}C_W\h_{\scriptscriptstyle W}
+C^\al\partial_\al \h_{\scriptscriptstyle
W}+\frac{1}{2}C_L\r_5\h_{\scriptscriptstyle W}
-i\r^\al\w(\partial_\al C_W-\overline\h_{\scriptscriptstyle W}\chi_\al)
+\w F_W ~, \nn \\
&& \de F_W=C^\al\partial_\al F_W-i\overline\w\r^\al[
\nabla_\al\h_{\scriptscriptstyle W}+i\r^\be\chi_\al
(\partial_\be C_W-\overline\chi_\be\h_{\scriptscriptstyle W})
-\chi_\al F_W] ~. \label{cbrsttr}
\end{eqnarray}
In terms of the new variables introduced in Sections 2 and 3 these
transformations can be rewritten as
\begin{eqnarray}
\de X&=&C^\al\dl_\al X-i(\w_-\psi_+-\w_+\psi_-) ~, \nn \\
\de\psi_\pm&=&C^\al\dl_\al\psi_\pm+\sqrt{\frac{2}{\lam^++\lam^-}}f_X\w_\pm
+{1\over2}(C^{1\prime}\pm\lam^\pm C^{0\prime})\psi_\pm \nn \\
&&\pm{2\w_\mp\over\lam^++\lam^-}\{\dot X\pm\lam^\mp X'
+i(\n_-\psi_+-\n_+\psi_-)\} ~, \nn \\
\de f_X&=&C^\al\dl_\al f_X+\frac{1}{2}\dl_\al C^\al f_X\nn \\
&& -i\sqrt{\frac{2}{\lam^++\lam^-}}\w_+\Bigl[\dot\psi_+-\lam^+\psi_+'
-\frac{1}{2}\lam^+{}'\psi_+-\frac{2\nu_-}{\lam^++\lam^-}(\dot X+\lam^- X'
-i\nu_+\psi_-)\Bigr] \nn \\
&& -i\sqrt{\frac{2}{\lam^++\lam^-}}\w_-\Bigl[\dot\psi_-+\lam^-\psi_-'
+\frac{1}{2}\lam^-{}'\psi_-+\frac{2\nu_+}{\lam^++\lam^-}(\dot X-\lam^+X'
+i\nu_-\psi_+)\Bigr] ~, \nn \\
\de \f&=&C_W+C^\al\dl_\al \f-i(\w_-\h_+-\w_+\h_-) ~, \nn \\
\de \h_\pm&=&-\h_{{\scriptscriptstyle W}\pm}+C^\al\dl_\al\h_\pm
+\sqrt{\frac{2}{\lam^++\lam^-}}\w_\pm f_L
+{1\over2}(C^{1\prime}\pm\lam^\pm C^{0\prime})\h_\pm \nn \\
&&\pm{2\w_\mp\over\lam^++\lam^-}\{\dot\f\pm\lam^\mp\f'
+i(\n_-\h_+-\n_+\h_-)\}~, \nn \\
\de f_L&=&f_W+C^\al\dl_\al f_L+\frac{1}{2}\dl_\al C^\al f_L\nn \\
&& -i\sqrt{\frac{2}{\lam^++\lam^-}}\w_+\Bigl[\dot\h_+-\lam^+\h_+'
-\frac{1}{2}\lam^+{}'\h_+-\frac{2\nu_-}{\lam^++\lam^-}(\dot \f+\lam^- \f'
-i\nu_+\h_-)\Bigr] \nn \\
&& -i\sqrt{\frac{2}{\lam^++\lam^-}}\w_-\Bigl[\dot\h_-+\lam^-\h_-'
+\frac{1}{2}\lam^-{}'\h_-+\frac{2\nu_+}{\lam^++\lam^-}(\dot \f-\lam^+\f'
+i\nu_-\h_+)\Bigr] ~, \nn \\
\de\lam^\pm&=&C^\al\dl_\al\lam^\pm\pm(\dot C^1\pm\lam^\pm\dot C^0)
-\lam^\pm(C^{1\prime}\pm\lam^\pm C^{0\prime})-4i\w_\mp\n_\mp ~, \nn \\
\de\x&=&C_W+C^\al\dl_\al\x+2C^{1\prime}+C^0{}'(\lam^+-\lam^-)
-i(\w_-\Lam_+-\w_+\Lam_-) ~, \nn \\
\de \ve&=&C_L+C^\al\dl_\al\ve+\frac{1}{2}(\lam^++\lam^-)C^0{}'
-\frac{i}{2}(\w_-\Lam_++\w_+\Lam_-) ~,\nn \\
\de\n_\pm&=&C^\al\dl_\al\n_\pm+(\dot C^0\pm\lam^\mp C^{0\prime})\n_\pm
-{1\over2}(C^{1\prime}\mp\lam^\mp C^{0\prime})\n_\pm
+\dot\w_\pm\pm\lam^\mp\w'_\pm\mp{1\over2}\lam^{\mp\prime}\w_\pm ~, \nn \\
\de\Lam_\pm&=&-\eta_{{\scriptscriptstyle W}\pm}+C^\al\dl_\al\Lam_\pm
+\sqrt{\frac{2}{\lam^++\lam^-}}\w_\pm f_G
+4C^0{}'\n_\mp
+{1\over2}(C^1{}'\pm\lam^\pm C^0{}')\Lam_\pm \nn \\
&&\pm{2\w_\mp\over\lam^++\lam^-}
\{\dot\x\pm\lam^\mp\x'-(\lam^+-\lam^-)'
-i(\n_+\Lam_--\n_-\Lam_+)\}+4\w_\mp' ~, \nn \\
\de f_G&=&f_W+C^\al\dl_\al f_G+\frac{1}{2}\dl_\al C^\al f_G \nn \\
&& -i\sqrt{\frac{2}{\lam^++\lam^-}}\w_+\Bigl[\dot \Lam_+-\lam^+ \Lam_+'
-\frac{1}{2}\lam^+{}' \Lam_+-4\nu_-' \nn \\
&& -\frac{2\nu_-}{\lam^++\lam^-}\{\dot \x+\lam^- \x'-(\lam^+-\lam^-)'
-i\nu_+ \Lam_-\}\Bigr] \nn \\
&& -i\sqrt{\frac{2}{\lam^++\lam^-}}\w_-\Bigl[\dot \Lam_-+\lam^- \Lam_-'
+\frac{1}{2}\lam^-{}' \Lam_--4\nu_+' \nn \\
&& +\frac{2\nu_+}{\lam^++\lam^-}\{\dot \x-\lam^+\x'-(\lam^+-\lam^-)'
+i\nu_- \Lam_+\}\Bigr] ~, \nn \\
\de C^0&=&C^\al\dl_\al C^0+{2i\over\lam^++\lam^-}(\w_+^2+\w_-^2) ~, \nn \\
\de C^1&=&C^\al\dl_\al C^1-{2i\over\lam^++\lam^-}(\lam^+\w_+^2-\lam^-\w_-^2)
{}~,\nn \\
\de\w_\pm&=&C^\al\dl_\al\w_\pm-{1\over2}(C^{1\prime}\mp\lam^\mp C^{0\prime})
\w_\pm-{2i\n_\pm\over\lam^++\lam^-}(\w_+^2+\w_-^2) ~, \nn \\
\de C_L&=&C^\al\dl_\al C_L+i\sqrt{\frac{2}{\lam^++\lam^-}}\w_-\w_+f_G \nn \\
&&+\frac{2i\w_+^2}{\lam^++\lam^-}
\Bigl[-\dot\ve+\lam^+\ve'+\lam^+{}'-i\nu_-\Lam_+
-\frac{1}{2}(\dot\x-\lam^+\x')\Bigr]
-\frac{i}{2}\w_+\h_{{\scriptscriptstyle W}-} \nn \\
&&+\frac{2i\w_-^2}{\lam^++\lam^-}
\Bigl[-\dot\ve-\lam^-\ve'+\lam^-{}'-i\nu_+\Lam_-
+\frac{1}{2}(\dot\x+\lam^-\x')
\Bigr]-\frac{i}{2}\w_-\h_{{\scriptscriptstyle W}+} ~, \nn \\
\de C_W&=&C^\al\dl_\al C_W
-i(\w_-\eta_{{\scriptscriptstyle W}+}-\w_+\eta_{{\scriptscriptstyle W}-}) ~,
\nn \\
\de\eta_{{\scriptscriptstyle W}\pm}&=&C^\al\dl_\al\eta_{{\scriptscriptstyle
W}\pm}
+{1\over2}(C^1{}'\pm\lam^\pm C^0{}')\eta_{{\scriptscriptstyle W}\pm}
+\sqrt{\frac{2}{\lam^++\lam^-}}\w_\pm f_W \nn \\
&&\pm{2\w_\mp\over\lam^++\lam^-}\{\dot C_W\pm\lam^\mp C_W'
+i(\n_-\eta_{{\scriptscriptstyle W}+}-\n_+\eta_{{\scriptscriptstyle W}-})\} \nn
\\
\de f_W&=&C^\al\dl_\al f_W+\frac{1}{2}\dl_\al C^\al f_W \nn \\
&& -i\sqrt{\frac{2}{\lam^++\lam^-}}\w_+\Bigl[
\dot\h_{{\scriptscriptstyle W}+}-\lam^+\h_{{\scriptscriptstyle W}+}'
-\frac{1}{2}\lam^+{}'\h_{{\scriptscriptstyle W}+}
-\frac{2\n_-}{\lam^++\lam^-}(\dot C_W+\lam^- C_W'-i\n_+
\h_{{\scriptscriptstyle W}-})\Bigr]~, \nn \\
&& -i\sqrt{\frac{2}{\lam^++\lam^-}}\w_-\Bigl[
\dot\h_{{\scriptscriptstyle W}-}+\lam^-\h_{{\scriptscriptstyle W}-}'
+\frac{1}{2}\lam^-{}'\h_{{\scriptscriptstyle W}-}
+\frac{2\n_-}{\lam^++\lam^-}(\dot C_W-\lam^+C_W'+i\n_-
\h_{{\scriptscriptstyle W}+})\Bigr]~, \nn \\
&& \label{ncbrsttr}
\end{eqnarray}
where $f_{X,L,G,W}$ are defined by
\begin{equation}
f_{X,L,G,W}=\sqrt{e}F_{X,L,G,W} ~. \label{ncauxf}
\end{equation}
It is straightforward to ascertain the nilpotency of the BRST
transformations.

\end{document}